\documentclass[a4paper,10pt]{article}
\usepackage[utf8]{inputenc}

\usepackage[max2]{authblk}

\usepackage{hyperref}

\usepackage{cite}

\usepackage{amsmath}
\usepackage{amssymb}

\usepackage{blkarray}
\usepackage{fullpage}

\usepackage{multirow}

\usepackage{graphicx}
\usepackage{subfigure}

\usepackage{xcolor}

\xdefinecolor{darkblue}{rgb}{0.,0.,0.7}
\xdefinecolor{darkgreen}{rgb}{0,0.43,0}
\xdefinecolor{darkorange}{rgb}{0.8,0.44,0}

\title{\begin{flushright}\large
\vspace{-1cm} DESY 14-041\\
 IFT-UAM/CSIC-14-058
\end{flushright}\vspace{1cm}\textbf{Neutralinos betray their singlino nature at the ILC}}
\author[1,2]{Gudrid Moortgat-Pick\footnote{\texttt{gudrid.moortgat-pick@desy.de}}}
\author[1]{Stefano Porto\footnote{\texttt{stefano.porto@desy.de}}}
\author[3]{Krzysztof Rolbiecki\footnote{\texttt{krzysztof.rolbiecki@desy.de}}}
\affil[1]{ \small II. Institut f\"{u}r Theoretische Physik, University of Hamburg, Germany\normalsize}
\affil[2]{\small DESY, Hamburg, Germany\normalsize}
\affil[3]{ \small  Instituto de Fisica Teorica IFT-UAM/CSIC, Madrid, Spain\normalsize}

\begin{document}

\maketitle

\begin{abstract}
It is one of the most challenging tasks at the Large Hadron Collider
 and at a future Linear Collider not only to observe physics
 beyond the Standard Model, but to clearly identify
 the underlying new physics model.  In this paper we concentrate on the
 distinction between two different supersymmetric models, the MSSM and
 the NMSSM, as they can lead to similar low energy spectra.  The NMSSM
 adds a singlet superfield to the MSSM particle spectrum
 and simplifies embedding a SM-like Higgs candidate with the measured
 mass of about 125.5~GeV.  
In parts of the parameter space the Higgs sector itself
 does not provide sufficient indications for the underlying model.  We show that
 exploring the gaugino/higgsino sectors could provide a meaningful way to distinguish the two models. Assuming that only
 the lightest chargino and neutralino masses and polarized cross
 sections $e^+e^-\to \tilde{\chi}^0_i\tilde{\chi}^0_j$,
 $\tilde{\chi}^+_i\tilde{\chi}^-_j$ are accessible at the linear collider, we
 reconstruct the fundamental MSSM parameters $M_1$, $M_2$, $\mu$,
 $\tan\beta$ and study whether a unique model distinction is
 possible based on this restricted information. Depending on the singlino admixture in the lightest
 neutralino states, as well as their higgsino or gaugino nature, we
 define several classes of scenarios and study the prospects of experimental differentiation.
\end{abstract}

\section{Introduction}

The discovery of a neutral scalar particle~\cite{Aad:2012tfa,Chatrchyan:2012ufa} with mass $\sim 125.5$ GeV at
the Large Hadron Collider (LHC) \cite{Chatrchyan:2013mxa,Aad:2014aba} has opened a plethora of
discussions about its identity.  While the experimental 
uncertainty suggests
the new particle to be the Standard Model Higgs boson, more data are
still needed to precisely determine its branching ratios, the CP
properties and the underlying model. The present results are in fact compatible with one of the
most promising Beyond the Standard Model candidates: supersymmetry
(SUSY)~\cite{Martin:1997ns}. The latter solves --- contrary to the Standard Model (SM) ---
the electroweak hierarchy puzzle, offers a
dark matter candidate and is consistent with grand unification of the
gauge couplings.

The most studied supersymmetric models are the Minimal Supersymmetric
Standard Model (MSSM)~\cite{Martin:1997ns} and its minimal extension,
the Next-to-Minimal Supersymmetric Standard Model (NMSSM)
\cite{Ellwanger:2009dp}. The NMSSM introduces a gauge singlet chiral
supermultiplet $\tilde{S}$ that allows for a relaxation of the
electroweak fine tuning conditions, compared to the MSSM.  On the other
hand, the --- so far --- negative result of LHC searches for
physics beyond the Standard Model
(BSM)~\cite{ATLAS-CONF-2014-006,CMS-PAS-SUS-13-020} does not favor any
of these models \textit{ a priori}.

In case of SUSY discovery at the LHC and/or at a linear
collider (LC) it is therefore important to understand how to entail the
underlying supersymmetric model, in particular how to distinguish
between NMSSM and MSSM. These two models have indeed a very similar
particle spectrum, with the exception for the superfield $\hat{S}$ in the NMSSM that
results in three additional states with respect to the MSSM: a CP-even
Higgs, a CP-odd Higgs and a fifth neutralino.

It is therefore well-motivated to look at the Higgs sector, where the
experiments are expected to give the most precise indications~\cite{Gupta:2012fy,Asner:2013psa}, and to complement
the information by studying
the (extended) neutralino sector of the NMSSM 
to look for deviations with respect to the MSSM.

Concerning the gaugino/higgsino sector, it has been shown that detecting the lightest chargino
$\tilde{\chi}^{\pm}_1$~\cite{Choi:2000ta},
and neutralino states $\tilde{\chi}^0_1$, $\tilde{\chi}^0_2$~\cite{Choi:2001ww,Choi:2002mc}, 
a full reconstruction of the MSSM chargino and
neutralino sectors through $\chi^2$-fits~\cite{Desch:2003vw,Desch:2006xp}, is possible
based on measuring the masses and their polarized cross sections.  
A fit disfavouring the MSSM suggests to look at minimal
extensions that modify the neutralino/chargino sector, \textit{in
primis} the NMSSM~\cite{MoortgatPick:2005vs,Porto:2014fca}.

In fact, the singlino admixtures of neutralino lightest states as well
as the higgsino and gaugino components of $\tilde{\chi}_1^0$ allow to
identify several classes of NMSSM scenarios.  Scenarios where the
singlino component in the light neutralinos
is significant (light
singlino scenarios), are often treated in the literature, featuring
production cross sections and phenomenology different from the
MSSM and are in principle easier to spot.  If the singlino, however, is heavy and
mainly present in $\tilde{\chi}_3^0$, $\tilde{\chi}_4^0$ or $\tilde{\chi}_5^0$,
the phenomenology is more MSSM-like and we distinguish the cases
where the main component of the lightest $\tilde{\chi}_1^0$
is higgsino-like (light higgsino scenarios) or gaugino-like (light
gaugino scenarios). Having a decoupled
singlino may result in a scenario that is experimentally not distinguishable
from the MSSM without further information about the heavier
neutralino states and the Higgs sector. Our analysis confirms these hints, concluding that a light and accessible singlino is one of the most efficient ways for model distinction together with a light singlet scalar.

The paper is organized in the following way:
first we introduce our proposed strategy to discriminate the different models in
section~\ref{Sec_Strategy} and describe 
 the classes of scenarios in section~\ref{Sec:Scenarios}. 
In that section we also try 
to clearly classify in which cases a unique distinction
between both models is possible based only on the light electroweak states
and to work out which further information is required in cases where the
light sector alone does not provide sufficient information for a model discrimination. 
Therefore we
perform scans in the $(\lambda,\,\kappa)$-plane, applying the most
recent phenomenological and experimental constraints from colliders, including
also dark matter experiments, and determine where the singlino admixtures are
such that the NMSSM cannot be misinterpreted as MSSM.  We summarize our
results in~\ref{Sec_Conclusions} and list details and parameters on the models
in the appendices~\ref{App:Sectors}, \ref{app-c}.


\section{Strategy \label{Sec_Strategy}}
As explained in the Introduction, the NMSSM adds to the MSSM an additional gauge singlet superfield $\hat{S}$ in
the Higgs sector. The most studied version of the NMSSM has a Lagrangian
with an accidental $\mathbb{Z}_3$ symmetry, obtained from the scale
invariant superpotential \cite{Ellwanger:2009dp}
\begin{equation}
 W_{\tiny\mathbb{Z}_3\mbox{-NMSSM}}\supset\lambda\,\hat{S}\hat{H}_u\cdot\hat{H}_d+\frac{\kappa}{3}\,\hat{S}^3\,.\label{Z3Superpotential}
\end{equation} 
$\hat{S}$ consists of a scalar Higgs singlet $S$ and the singlino
$\tilde{S}$. The additional dimensional parameters $A_{\lambda}$ and $A_{\kappa}$ appear in the Higgs sector soft terms:
\begin{equation}
 \mathcal{L}_{\tiny \rm soft, \mathbb{Z}_3\mbox{-NMSSM}}\supset -\lambda A_{\lambda}H_u\cdot H_d S-\frac{1}{3}A_{\kappa}S^3.\label{Z3Soft}
\end{equation}
The singlet $S$, see Eqs.~\eqref{Z3Superpotential} and \eqref{Z3Soft}, mixes
due to the electroweak symmetry breaking
with the MSSM Higgs doublets $H_u, \, H_d$, resulting in three CP-even
neutral scalars $h_1,\,h_2,\,h_3$ and two CP-odd neutral scalars
$a_1,\,a_2$. Correspondingly, the singlino $\tilde{S}$ mixes  
with the higgsinos and the gauginos, resulting in five neutralino mass eigenstates. 
Therefore, determining the nature of weakly coupling scalars or neutralinos is the first way to discriminate between NMSSM and
MSSM.  

In the light of the expected high accuracy in the Higgs sector
measurements~\cite{Asner:2013psa}, it is a common practice to compare MSSM
and NMSSM scenarios looking at the Higgs sector, in particular at the
Higgs decays~\cite{Benbrik:2012rm,Beskidt:2013gia,Pandita:2014nsa}.  The
case in which a very light CP-even and/or a light CP-odd scalars have high
singlet component and allow new decay channels for the SM-like Higgs
scalar affecting its decay width and branching ratios has been
explored~\cite{Ellwanger:2003jt}. On the same footing, looking at the
extended NMSSM neutralino sector is very well motivated, especially for
linear collider phenomenology, due to the high precision in the
electroweak sector. This can be crucial in case of
relatively heavy singlet states in comparison with the SM-like Higgs,
such that the observed Higgs sector can be interpreted within both the
MSSM and the NMSSM. 
In such scenarios with heavy decoupled 
states, the corresponding signatures at the LHC
would indeed be very similar in both models~\cite{Asner:2013psa}.

We are therefore interested to understand how much information can be obtained from the neutralino and chargino sector for the model distinction. In the MSSM, the parameters $M_1$, $M_2$, $\mu$, $\tan \beta$ fully describe the chargino and neutralino sector. One should note that these are fundamental parameters without any assumption on the SUSY breaking scheme.
 Precise determination of
 these parameters is possible at a linear collider~\cite{Choi:2000ta,Choi:2001ww,Choi:2002mc,Desch:2003vw,Bharucha:2012ya}, provided that 
 $\tilde{\chi}^0_1$, $\tilde{\chi}^0_2$ and $\tilde{\chi}^{\pm}_1$
 can be produced at the LC and their masses as well as the polarized
 cross sections
 $\sigma(e^+e^-\rightarrow\tilde{\chi}^0_1\tilde{\chi}^0_2)$,
 $\sigma(e^+e^-\rightarrow\tilde{\chi}^+_1\tilde{\chi}^-_1)$ are
 measured.  
An accurate and rather model-independent determination of $M_1$, $M_2$, $\mu$, 
$\tan \beta$ is performed by a $\chi^2$-minimisation that selects
 the parameters fitting the experimental results. Such analysis can be strengthened if the
 mass of the heavier neutralino states can be inferred from combined
 analyses of LHC and LC data~\cite{Desch:2003vw}.
 
The possibility of reconstructing the MSSM chargino-neutralino sector
parameters can then be exploited as a tool for the distinction between
the MSSM and the NMSSM~\cite{MoortgatPick:2005vs}.  Given experimental
observation of $\tilde{\chi}^0_1$, $\tilde{\chi}^0_2$ and
$\tilde{\chi}^{\pm}_1$, a result of the $\chi^2$-fit that excludes the
MSSM at 95\% confidence level (C.L.), may suggest the NMSSM.  It has indeed been shown~\cite{MoortgatPick:2005vs} that relatively different mixing for MSSM and NMSSM
 scenarios can lead to very similar neutralino and chargino mass
spectra in both models; this
is of course also true in case of a scenarios with similar soft parameters
and a decoupled singlet superfield.

\begin{figure}[htb]\centering
\includegraphics[width=1\textwidth]{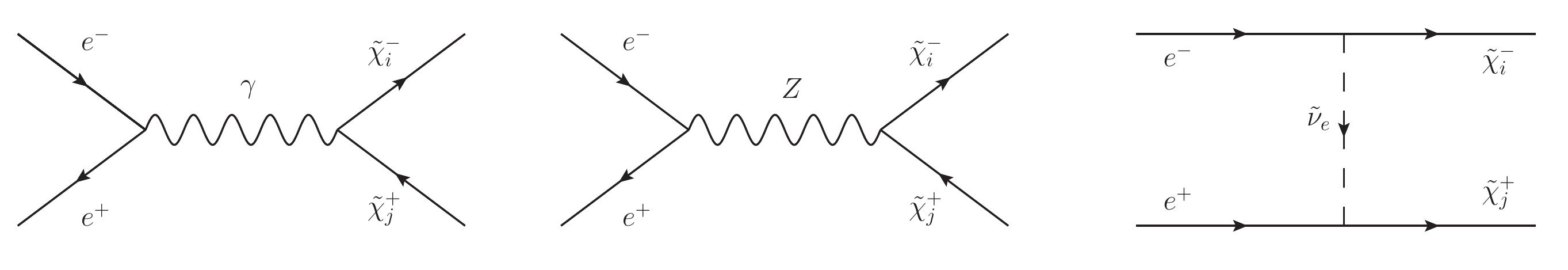}
\caption{Chargino tree-level production channels at $e^+e^-$ colliders.\label{CharJD}}
\end{figure}

\begin{figure}[htb]\centering
\includegraphics[width=1\textwidth]{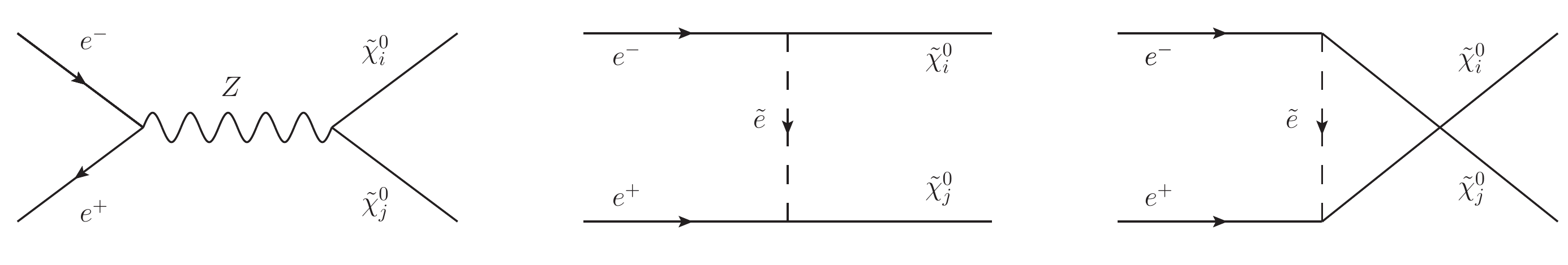}
\caption{Neutralino tree-level production channels at $e^+e^-$ colliders.\label{NeuJD}}
\end{figure}

Following this idea, we outline our strategy:

\begin{itemize}
 \item \textbf{Scenario selection.} We identify NMSSM scenarios that
       present a mass spectrum for $\tilde{\chi}^{\pm}_1$,
       $\tilde{\chi}^{0}_1$, $\tilde{\chi}^{0}_2$ and low Higgs spectrum
       that can be attributed also to a MSSM scenario. We calculate the
       corresponding NMSSM neutralino and chargino tree-level masses and
       polarized cross-sections for the processes
       $e^+e^-\rightarrow\tilde{\chi}^+_1\tilde{\chi}^-_1$ and
       $e^+e^-\rightarrow\tilde{\chi}^0_1\tilde{\chi}^0_2$
cf. (Figures~\ref{CharJD} and~\ref{NeuJD}). 
 
\item \textbf{Constraints.} Each scenario has to
fulfill a series of phenomenological and experimental constraints
implemented in \texttt{NMSSMTools-4.2.1}, that includes
\texttt{NMHDECAY}~\cite{Ellwanger:2004xm,Ellwanger:2005dv,Belanger:2005kh} and
\texttt{NMSDECAY}~\cite{Das:2011dg,Muhlleitner:2003vg}. These tools
calculate the Higgs sector parameters, SUSY particle masses at the loop level and their decays,
and confront them with limits from
LEP, LHC and EW precision
constraints. An interface to \texttt{MicrOMEGAS}~\cite{Belanger:2013oya}
       provides dark matter constraints, including the latest LUX~\cite{Akerib:2013tjd} and
Planck~\cite{Ade:2013zuv} results. The LSP relic density is required to be $\Omega_{\rm
       LSP}h^2< 0.131$, where $h$ is the Hubble constant in units of 100
       km/(s$\cdot$Mpc). Higgs sector constraints are further controlled
       using \texttt{HiggsBounds-4.0.0}~\cite{Bechtle:2013wla} and
       \texttt{HiggsSignals-1.0.0}~\cite{Bechtle:2013xfa}, such that a
       scenario is accepted only if
      compatible with current data at the 95\% (C.L.).  

 \item \textbf{Experimental assumption.} We assume, for each NMSSM scenario, an observation of
       $\tilde{\chi}^{\pm}_1$, $\tilde{\chi}^{0}_1$
       and $\tilde{\chi}^{0}_2$ at the ILC together with their total cross sections
       $\sigma(e^+e^-\rightarrow\tilde{\chi}^+_1\tilde{\chi}^-_1)$,
       $\sigma(e^+e^-\rightarrow\tilde{\chi}^0_1\tilde{\chi}^0_2)$ with
       electron-positron beam polarizations
       $(\mathcal{P}_{e^{-}},\mathcal{P}_{e^{+}})= (\pm0.9,\mp0.55)$ at
       $\sqrt{s}=350$ GeV (corresponding to the $t\bar{t}$-threshold)
       and at $\sqrt{s}=500$ GeV. A precision of 0.5\% on
       the masses and 1\% on the cross
sections is assumed~\cite{AguilarSaavedra:2001rg,Baer:2013cma}. If
       kinematically accessible, also $m_{\tilde{\chi}^0_3}$, and the processes
       $e^+e^-\rightarrow\tilde{\chi}^0_1\tilde{\chi}^0_3$,
       $e^+e^-\rightarrow\tilde{\chi}^0_2\tilde{\chi}^0_3$
       are considered. 
 
\item $\mathbf{\chi^2}$\textbf{-fit to MSSM.} The measured quantities
      and errors are used to perform a MSSM parameter determination
      through the $\chi^2$-fit following the recipe in 
\cite{Desch:2003vw}, similarly
      to~\cite{MoortgatPick:2005vs}.  We 
apply 
the $\chi^2$-fit using \texttt{Minuit} \cite{James1975343}, that minimizes the $\chi^2$ function defined as 
\begin{equation}
     \chi^2=\sum_i\left|\frac{\mathcal{O}_i-\bar{\mathcal{O}}_i}{\delta\mathcal{O}_i}\right|^2\mbox{,}
\end{equation}
where $\mathcal{O}_i$ are the input observables, $\delta O_i$ are the associated experimental uncertainties and $\bar{O}_i$ are the theoretical values of the observables calculated using the fitted MSSM parameters.
The unknowns of the fit will be $M_1,\,M_2,\,\mu,\,\tan\beta$ and
      $m_{\tilde{\nu}_e}$.\footnote{The mass $m_{\tilde{\nu}_e}$ is
      related to the selectron masses by applying the SU(2)
      relation $m_{\tilde{\nu}_e}^2=m_{\tilde{e}_L}^2+\cos(2\beta)\cos^2\theta_Wm_Z^2$ and $m_{\tilde{e}_L}=m_{\tilde{e}_R}$.}
In the case of high $\tan \beta$, its extraction could be difficult, and
       only a lower limit could be set.
A fit that is not consistent with the MSSM 95\% C.L., may give hints towards the
       NMSSM and model distinction. If this is not the case, more
       information is needed to be included to establish the nature of
       the observed model. The limiting (95\% C.L.) value of $\chi^2$ varies  for different scenarios under consideration depending on the number of observables used in the fit.

\item \textbf{Information from the Higgs sector.} 
If the singlet is relatively light and has a substantial mixing with the
      SM-like Higgs, one could observe deviations from the SM
      predictions that cannot be accommodated within the MSSM at the
      same time. In our case, we expect small departure
      from the SM values and we limit ourselves to comparing the NMSSM
      predictions to the SM model by doing a $\chi^2$-fit of the reduced
      couplings of the SM-like Higgs to $g,\gamma,W,Z,b,c,\tau$. If the couplings do not
      differ too much from the SM, such a scenario could always be
      accommodated within the MSSM in the decoupling limit as
      well. Alternatively, one could consider a possibility of detection additional singlet-like states, but this analysis is beyond the scope of the current paper.

\end{itemize}

\section{Classes of scenarios}\label{Sec:Scenarios}
The singlino ($\tilde{S}$) admixtures of the lightest neutralino states
$\tilde{\chi}^0_1$ and $\tilde{\chi}^0_2$ suggest the classification of
NMSSM scenarios with the following limiting cases:
\begin{enumerate}
 \item \textbf{Light singlino (LS) scenarios:} high $\tilde{S}$ admixture in the light states $\tilde{\chi}^{0}_{1}$ or $\tilde{\chi}^{0}_{2}$.
 
\item \textbf{Light higgsino (LH) scenarios:} higgsino-like $\tilde{\chi}^{0}_{1}$, with $\mu_{\rm eff}<M_1,M_2$ and high $\tilde{S}$ admixture mainly in 
       $\tilde{\chi}^{0}_{3},\tilde{\chi}^{0}_{4},\tilde{\chi}^{0}_{5}$.
       
\item \textbf{Light gaugino (LG) scenarios: }gaugino-like $\tilde{\chi}^{0}_{1}$, with $\mu_{\rm eff}>M_1,M_2$ and high $\tilde{S}$ admixture mainly in
       $\tilde{\chi}^{0}_{3},\tilde{\chi}^{0}_{4},\tilde{\chi}^{0}_{5}$.
\end{enumerate}
Exploring these classes of scenarios allows to
embed also  the intermediate
cases of mixed lightest neutralino nature.

A high singlino admixture in $\tilde{\chi}^0_1$ and/or
       $\tilde{\chi}^0_2$ as in case 1 may signal beyond-MSSM physics. 
       A fit reconstructing the higgsino and gaugino components hypothesizing MSSM would give very different
       result with respect to the original NMSSM. In such a case, the outlined strategy for model distinction 
       seems promising, see~\cite{MoortgatPick:2005vs} and section~\ref{Subsection:LightSinglino}, as different gaugino and neutralino admixtures lead to modified cross sections, production channels, as well as decays.

       In cases 2 and 3, instead, both spectra and admixtures of the
       detected states $\tilde{\chi}^0_1$ and $\tilde{\chi}^0_2$ could
       result in a MSSM-like phenomenology, therefore it is likely that
       the fit is still compatible with the MSSM, see
       subsections~\ref{SubsecHIGLSP} and \ref{SubsecGAULSP}. In these
       cases one should ask how to efficiently integrate informations
from heavier neutralino states,
       and/or from the Higgs sector. 

Given a fixed $\mu_{\rm eff}=\lambda s$, the key parameters of the NMSSM
neutralino sector are $\lambda$ and $\kappa$ as they regulate the
singlino admixture in the mass eigenstates, see the NMSSM neutralino
mass matrix, Eq.~\eqref{NMSSMneutralinoMassMatrix} in
appendix~\ref{App:Sectors}.
In two heavy-singlino cases, see examples in
subsections~\ref{SubsecHIGLSP} and \ref{SubsecGAULSP}, we scan a grid of
ten thousand points in the $(\lambda,\,\kappa)$-plane for values
$\lambda\in[0,0.7]$ and $\kappa\in[0,0.7]$, to study how the model
discrimination method works at the ILC along the $(\lambda,\,\kappa)$-plane,
as the singlino admixtures vary.  For each point passing the previous
phenomenological and experimental constraints, we perform the
$\chi^2$-fit described above. These scans allow to see how the singlino
``mass'' vary along the $(\lambda,\,\kappa)$-plane, and to observe areas
in which the singlino is mostly very heavy and decoupled, areas in which
the singlino is placed among the lightest neutralino states, as well as
regions with mixed behaviour.

\subsection{Light singlino scenario\label{Subsection:LightSinglino}}
As a first example, we choose an NMSSM scenario with wino $\tilde{\chi}^0_1$ but with high singlino components in $\tilde{\chi}^0_2$ (and $\tilde{\chi}^0_3$). We refer to it as the light singlino scenario (LS). The lower neutralino/chargino spectrum can be reproduced by an MSSM scenario with different $M_1,\,M_2,\,\mu,\,\tan\beta$, see Table~\ref{Tabel:LSscenario}. Both for LS and the corresponding MSSM scenario we have $M_1>M_2$, as it is common in AMSB models.  We set $A_{\lambda}=4200$~GeV and $A_{\kappa}=-200$~GeV. For the remaining parameters of the NMSSM scenario, we refer to appendix \ref{Appendix:lightSinglino}. A SM-like Higgs with $m_h=125$~GeV is reproduced.

\begin{table}
 \begin{center}
\begin{tabular}{|c||c|c|c|c|c|c|}\hline
 &$M_1$ \footnotesize [GeV]\small&$M_2$ \footnotesize [GeV]\small&$\mu$,\,$\mu_{\rm eff}=\lambda\cdot x$ \footnotesize [GeV]\small&$\tan\beta$&$\lambda$&$\kappa$\\\hline
  MSSM&406&115.8&354&8&-&-\\\hline
 NMSSM&365&111&484&9.5&0.16&0.0585\\\hline
\end{tabular}             \end{center}
\caption{Neutralino and chargino parameters for the NMSSM scenario LS and for the corresponding MSSM scenario.\label{Tabel:LSscenario}}
\end{table}

The Higgs spectrum is given in Table~\ref{tab:higgsLS}.\footnote{In this study we used: $m_t$=173.07 GeV, $m_Z$=91.1876 GeV, $\Gamma_Z$=2.4952 GeV, $m_W$= 80.385, $\Gamma_W$=2.085 GeV, $\alpha_{\small\mbox{em}\normalsize}$= 1/127.92, $\alpha_{\small\mbox{s}\normalsize}(m_Z) $=0.1184, with $\sin^2\theta_W=1-m_W^2/m_Z^2$.} The mass $m_{{h}_1}$ can be easily reproduced within the corresponding
MSSM scenario with a proper choice of the stop soft parameters. The
states ${h}_2$ and ${a}_1$, being both $\sim100\%$ singlets, are not
expected to be visible both at the LHC and ILC because they are not
directly coupling to other particles and are relatively heavy. A
detailed analysis could point a way to observe these states but this is
beyond scope of this work.

\begin{table}
\small\begin{center}
\begin{tabular}{|c||c|c|c|c|c|c|}\hline
 &$m_{{h}_1}$ [GeV]&$m_{{h}_2}$ [GeV]&$m_{{h}_3}$ [GeV]&$m_{{a}_1}$ [GeV]&$m_{a_2}$ [GeV]&$m_{H^{\pm}}$ [GeV]\\\hline
NMSSM&124.9&303.0&4467.3&324.0&4467.3&4468.1\\\hline
\end{tabular}             \end{center}\normalsize
\caption{LS scenario: Higgs spectrum calculated at the 1-loop level with full
2-loops contributions from bottom/top Yukawa
 couplings
with \texttt{NMSSMTools}\cite{Ellwanger:2004xm,Ellwanger:2005dv,Belanger:2005kh}.\label{tab:higgsLS} }
\end{table}

\normalsize The tree-level masses for the neutralino/chargino sector are
listed in Table~\ref{tab:lhmasses}. The light part of the spectrum is nearly indistinguishable
between the two models, with $\tilde{\chi}^0_1\sim\tilde{W}$. However,
the other lighter states $\tilde{\chi}^0_2,\,\tilde{\chi}^0_3$ feature
different admixtures, see Table~\ref{mix-ls}, leading to different production cross sections and
relative importance of the production channels.

\begin{table}
\small\begin{center}
\begin{tabular}{|c||c|c|c|c|c|c|c|}\hline
 &$m_{\tilde{\chi}^0_1}$ [GeV]&$m_{\tilde{\chi}^0_2}$ [GeV]&$m_{\tilde{\chi}^0_3}$ [GeV]&$m_{\tilde{\chi}^0_4}$ [GeV]&$m_{\tilde{\chi}^0_5}$ [GeV]&$m_{\tilde{\chi}^{\pm}_1}$ [GeV]&$m_{\tilde{\chi}^{\pm}_2}$ [GeV]\\\hline
MSSM&104.8&350.4&360.1&426.7&-&105.1&375.0\\\hline
NMSSM&104.9&350.1&360.5&489.7&504.1&105.1&498.5\\\hline
\end{tabular}             \end{center}\normalsize
\caption{Neutralino and chargino masses in the LS scenario and in the corresponding reference MSSM scenario. The mass difference
$m_{\tilde{\chi}^{\pm}_1}-m_{\tilde{\chi}^0_1}$ receives significant
positive NLO corrections. Here, we only use tree-level masses, however
for such a quasi-degenerate states the mass measurement usually has a
larger uncertainty than the mass difference itself so in a more
realistic setting one should use the mass difference as an input
rather than the actual masses, see e.g.\ Ref.~\cite{Berggren:2013vfa}.\label{tab:lhmasses} }
\end{table}

\begin{table}
\small\begin{center}
\begin{tabular}{|c||c|c|}\hline
&MSSM&NMSSM\\\hline
 $\tilde{\chi}^0_1$&$\sim93\%\,\tilde{W}$&$\sim97\%\,\tilde{W}$\\\hline
  $\tilde{\chi}^0_2$&$\sim  26\% \,\tilde{B}+69\%\,\tilde{H}_{u,\,d}$&$\sim  22\% \,\tilde{B}+73\%\,\tilde{S}$\\\hline
   $\tilde{\chi}^0_3$&$\sim \tilde{H}_{u,\,d}$&$\sim  72\% \,\tilde{B}+25\%\,\tilde{S}$\\\hline
\end{tabular}             \end{center}
\normalsize
\caption{The dominant admixtures of the three lightest neutralinos in
 the LS scenario  and in the corresponding MSSM scenario.\label{mix-ls}}
\end{table}

We take $m_{\tilde{e}_L}=303.5$~GeV,  assuming
$m_{\tilde{e}_L}=m_{\tilde{e}_R}$ and
$m_{\tilde{\nu}_e}^2=m_{\tilde{e}_L}^2+\cos(2\beta)\cos^2\theta_Wm_Z^2$. 
The production cross sections are listed in Table~\ref{tab-lh-cross}. For the fit to the MSSM we only
include NMSSM cross sections larger than $1$~fb. The
relatively light NMSSM $\tilde{\chi}^0_3$ can be produced with a sizeable
cross section at 500~GeV, therefore we also include
$\sigma(e^+e^-\rightarrow\tilde{\chi}^0_1\tilde{\chi}^0_3)$ for
$P=(-0.9,0.55)$ in the fit.

\begin{table}
\begin{center}
\begin{tabular}{|c||c|c||c||c|c|}\hline
\multicolumn{6}{|c|}{  $\sigma(e^+e^-\rightarrow\tilde{\chi}^+_1\tilde{\chi}^-_1)$ \,\, [fb]}\\\hline
  $\sqrt{s}=$350 GeV&\scriptsize \textbf{MSSM} \normalsize & \scriptsize\textbf{ NMSSM} \normalsize&$\sqrt{s}= $500 GeV&\scriptsize\textbf{ MSSM} \normalsize & \scriptsize\textbf{ NMSSM }\normalsize\\\hline\hline
$P=(-0.9,0.55)$&2491.0&2575.3&$P=(-0.9,0.55)$&1165.4&1213.0
\\\hline
$P=(0.9,-0.55)$&39.5&42.4&$P=(0.9,-0.55)$&18.3&18.8
\\\cline{1-3}\hline
\end{tabular} \end{center}
\begin{center}
\begin{tabular}{|c||c|c||c||c|c|}\hline
\multicolumn{3}{|c||}{  $\sigma(e^+e^-\rightarrow\tilde{\chi}^0_1\tilde{\chi}^0_2)$ \,\, [fb]}&\multicolumn{3}{c|}{$\sigma(e^+e^-\rightarrow\tilde{\chi}^0_1\tilde{\chi}^0_3)$\,\, [fb]}\\\hline
  $\sqrt{s}=$500 GeV&\scriptsize \textbf{MSSM} \normalsize & \scriptsize\textbf{ NMSSM} \normalsize&$\sqrt{s}= $500 GeV&\scriptsize\textbf{ MSSM} \normalsize & \scriptsize\textbf{ NMSSM }\normalsize\\\hline\hline
$P=(-0.9,0.55)$&24.1&8.6&$P=(-0.9,0.55)$&25.1&15.0
\\\hline
$P=(0.9,-0.55)$&0.4&0.1&$P=(0.9,-0.55)$&5.7&0.2
\\\cline{1-3}\hline
\end{tabular} 
\end{center}
\caption{The production cross sections of
 $e^+e^-\to\tilde{\chi}^+_1\tilde{\chi}^-_1$,
 $\tilde{\chi}^0_1\tilde{\chi}^0_2$, $\tilde{\chi}^0_1\tilde{\chi}^0_3$
 in the LS scenario and the corresponding MSSM scenario at $\sqrt{s}=350$ and 500 GeV. \label{tab-lh-cross}}
\end{table}

The fitted MSSM parameters are then
\begin{align}
 &M_1=430.0\pm 1.6\mbox{ GeV,\,\,\, }\,\,M_2=111.8\pm0.8\mbox{ GeV, }\nonumber\\
 &\mu_{\rm eff}=370.4\pm0.7\mbox{ GeV, }\,\,\,m_{\nu_e}=310.6\pm2.8\mbox{ GeV}
\label{eq-mssm-fit}
\end{align}
and $\tan\beta$
 remains  unconstrained.
These parameters {would be consistent with
neutralino and chargino masses in the MSSM, listed in Table}~\ref{tab-mssm-fit}.

\begin{table}
\small\begin{center}
\begin{tabular}{|c||c|c|c|c|c|c|}\hline
 &$m_{\tilde{\chi}^0_1}$ [GeV]&$m_{\tilde{\chi}^0_2}$ [GeV]&$m_{\tilde{\chi}^0_3}$ [GeV]&$m_{\tilde{\chi}^0_4}$ [GeV]&$m_{\tilde{\chi}^{\pm}_1}$ [GeV]&$m_{\tilde{\chi}^{\pm}_2}$ [GeV]\\\hline
$\mbox{MSSM}_{\mbox{fit}}$&106.0&368.0&378.0 &445.9&106.1&389.1\\\hline
\end{tabular}             \end{center}\normalsize
\caption{MSSM neutralino and chargino masses based on the resulting
 parameters from the fit, see Eq.~\eqref{eq-mssm-fit}. \label{tab-mssm-fit}}
\end{table}

The fit with $10-5=5$ degrees of freedom (d.o.f.)
gives $\chi^2=62.6$,
{clearly} stating 
that the hypothesized model (MSSM) is not compatible with the
experimental data (with the 95\% C.L. being $\chi^2 < 11.1$). 
This could be additionally confirmed by the mass of
the heavy neutralino, $m_{\tilde{\chi}^0_4}$, if it is eventually measured at the higher center-of-mass energy. Additionally, we note that the predicted mass of the heavy chargino, 
$m_{\tilde{\chi}^\pm_2} = 389.1$~GeV, makes production of the mixed chargino pair, $\tilde{\chi}^\pm_1 \tilde{\chi}^\mp_2$ possible. The expected cross section, $\sim 3$~fb, could in principle allow for its measurement at $\sqrt{s} = 500$~GeV. The non-observation would provide another hint of the non-minimal nature of chargino/neutralino sector. A
non-minimal nature of the neutralino sector would be required to explain
the measurements with one of the possible candidates being
NMSSM. This first example
shows that an effective model distinction in the case of high admixture
of singlino in the lightest neutralino is possible exploiting the
outlined procedure.

\subsection{Light higgsino scenario, $\mu_{\rm eff}<M_1<M_2$\label{SubsecHIGLSP}}

We consider here an NMSSM scenario {with a light higgsino (LH), } whose chargino/neutralino parameters are: 
\begin{equation}
 M_1=450\mbox{ GeV,\,\,\, }\,\,\,M_2=1600\mbox{ GeV, }\,\,\,\mu_{\rm eff}=\lambda\,s=120\mbox{ GeV, }\,\,\,\tan\beta=27\,,\label{HigLSP_parameters}
\end{equation}
while we have $\lambda\in[0, 0.7]$ and $\kappa\in[0, 0.7]$ as described
above; $\mu_{\rm eff}$ is kept fixed by varying the singlet vacuum
expectation value (vev) $s $. The $\hat{S}$ soft parameters are $
A_{\lambda}=3000 $~GeV, $A_{\kappa}=- 30$~GeV. The first generation
sfermion masses, needed for the production cross sections, are {set to}
\begin{equation}\label{HigLSP_sfMass}
 m_{\tilde{e}_L}=303.5\mbox{ GeV},\,\,\, m_{\tilde{e}_L}=m_{\tilde{e}_R},\,\,\,m_{\tilde{\nu}_e}^2=m_{\tilde{e}_L}^2+\cos(2\beta)\cos^2\theta_Wm_Z^2,
\end{equation}
while squarks masses are $>1$ TeV. For the full set of soft parameters, we refer to Appendix~\ref{Appendix:lightHiggsino}.

\begin{figure}[t!]
\begin{center}
 \includegraphics[width=3.5in]{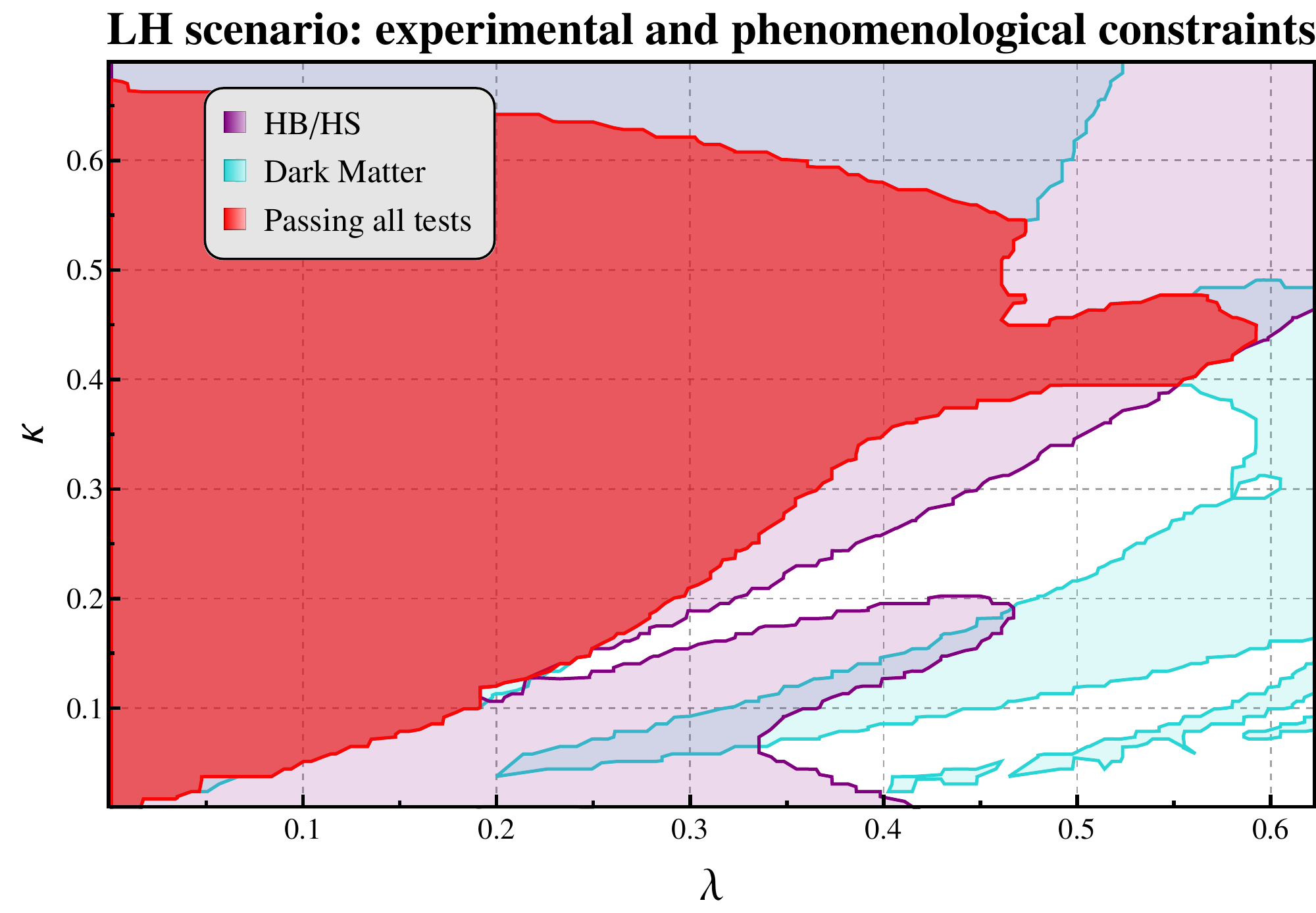}
 \caption{Light higgsino scenario: regions in the $(\lambda,\kappa)$-plane allowed by experimental and phenomenological constraints. The light-blue-shaded regions delimited by the light blue boundary pass dark matter constraints. The coloured regions delimited by the purple boundary pass checks within \texttt{HiggsBounds} and \texttt{HiggsSignals}.  The red area  is allowed by all the constraints.\label{HigLSP_Exclusions}}
\end{center}
\end{figure}

In Figure~\ref{HigLSP_Exclusions} we show the result of the scan in the
NMSSM $(\lambda,\kappa)$-plane after our tests. Light-blue-shaded area
corresponds to points that pass DM constraints;\footnote{Here and in the
following, we allow DM density to be below Planck~\cite{Ade:2013zuv} measured value.} the points within
purple-shaded boundary area pass the Higgs sector constraints from
\texttt{HiggsBounds} and \texttt{HiggsSignals}. The solid red area is
the region allowed by all the constraints, phenomenological and
experimental ones, implemented within \texttt{NMSSMTools},
\texttt{HiggsBounds} and \texttt{HiggsSignals}.

\begin{figure}[htb]\centering
\subfigure[]{\label{HIGLSPN1mass}\includegraphics[width=.49\textwidth]{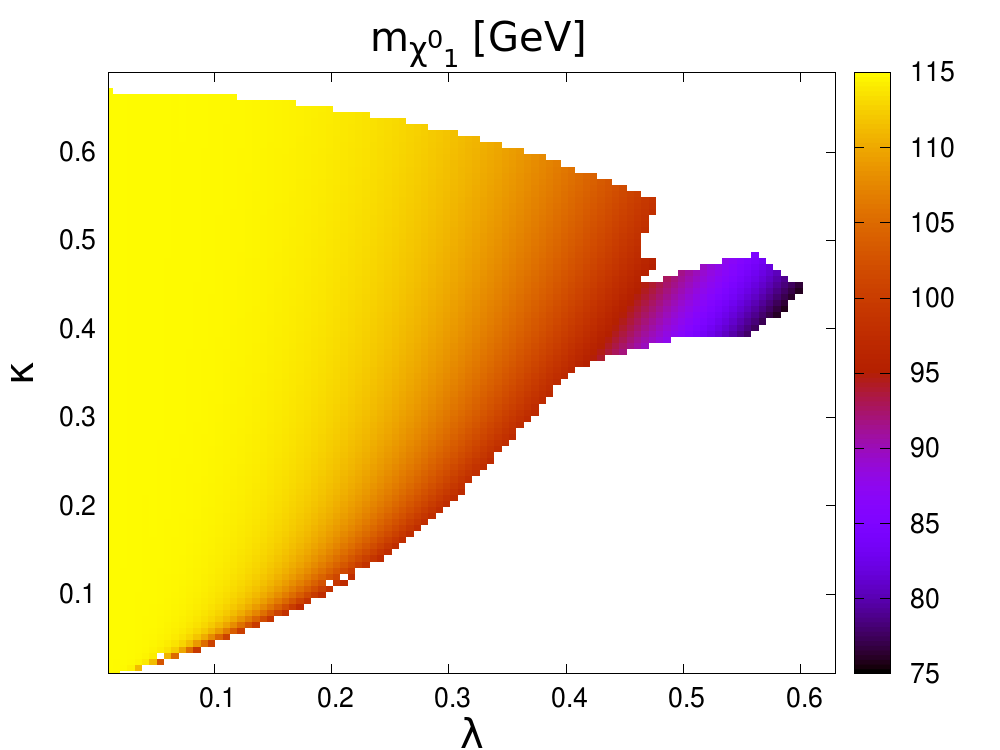}}
\subfigure[]{\label{HIGLSPN15}\includegraphics[width=.49\textwidth]{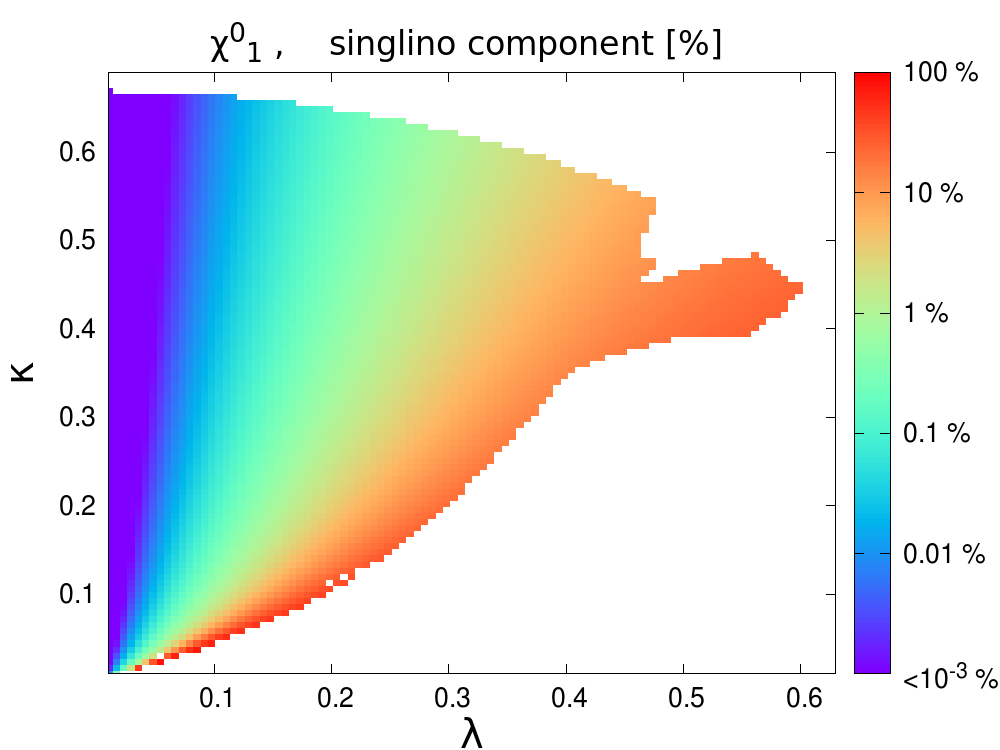}}
\caption{LH scenario: (a) the mass $m_{\tilde{\chi}^0_1}$, in GeV; (b) the $\tilde{S}$ component of $\tilde{\chi}^0_1$, in \%.\label{HIGN1plots}}
\end{figure}

\begin{figure}[htb]\centering
\subfigure[]{\label{HIGLSPN1N2_350M9P55}\includegraphics[width=.49\textwidth]{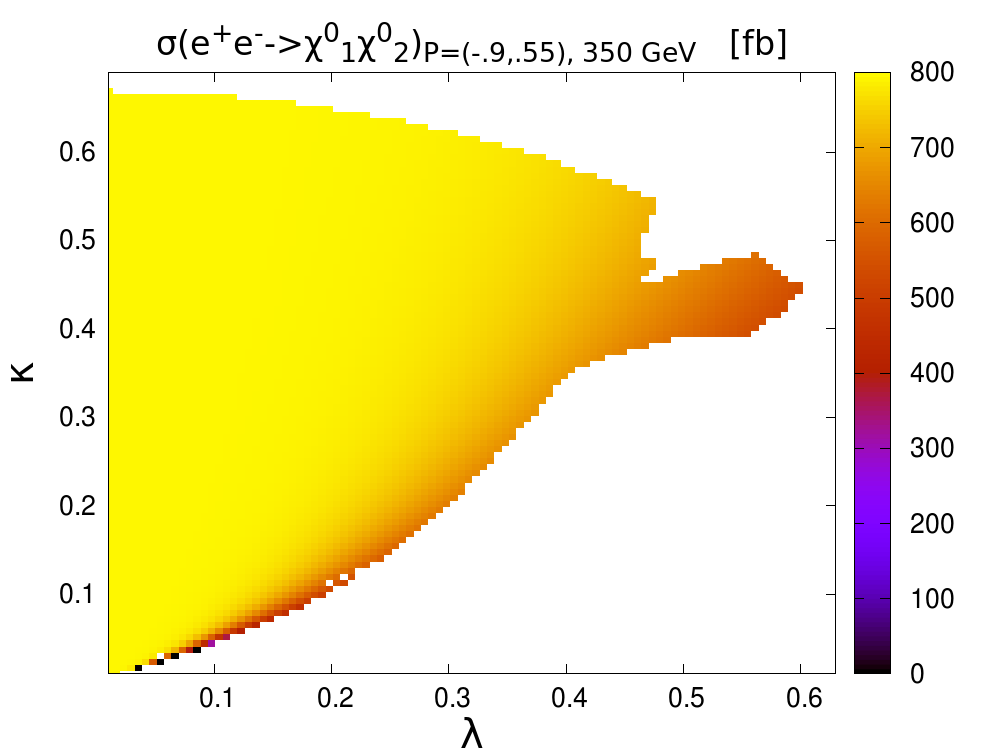}}
\subfigure[]{\label{HIGLSPN2N3_500M9P55}\includegraphics[width=.49\textwidth]{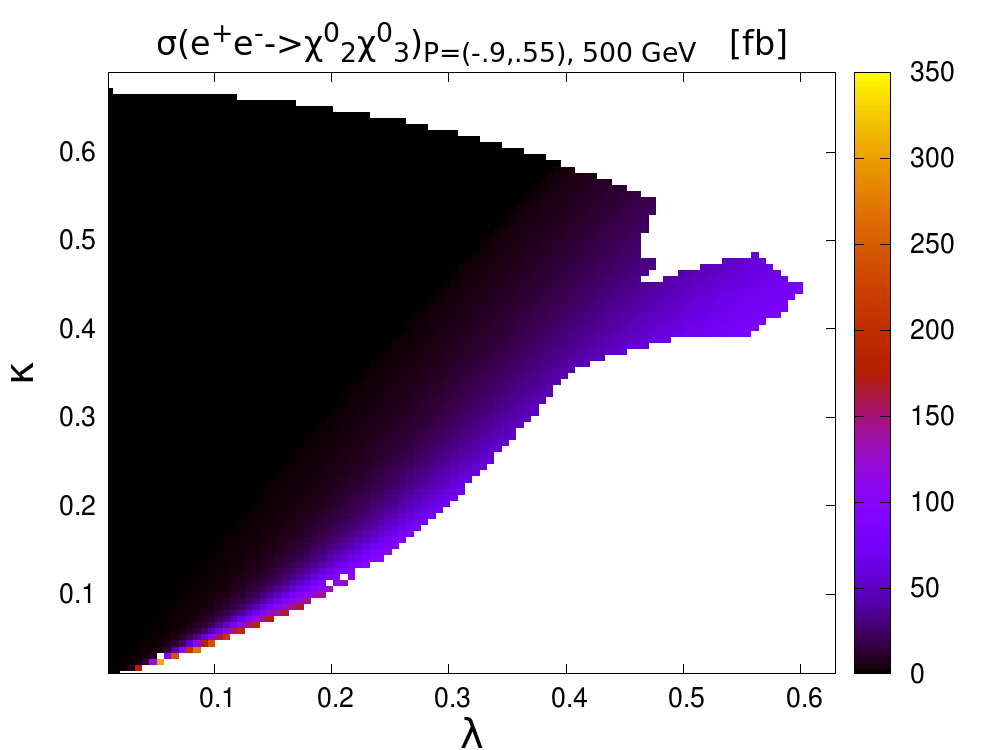}}
\caption{Production cross sections in the LH scenario: (a) $\sigma(e^+e^-\rightarrow\tilde{\chi}^0_1\tilde{\chi}^0_2)$ for $P= (-0.9,0.55)$ at $\sqrt{s} = 350$~GeV, in fb; (b) $\sigma(e^+e^-\rightarrow\tilde{\chi}^0_2\tilde{\chi}^0_3)$ for $P=(-0.9,+0.55)$ at $\sqrt{s} = 500$~GeV, in fb. \label{HIGN1N2_prod_plots}}
\end{figure}

As a reference MSSM scenario, we select the one with $M_1$,
$M_2$, $\mu=\mu_{\rm eff}$, $\tan\beta$ and the slepton masses given in
Eqs.~\eqref{HigLSP_parameters} and \eqref{HigLSP_sfMass}, to show that
the light neutralino spectrum  and 
production cross sections, {see Table}~\ref{HigLSPTab_MSSMXsecs},
may be very similar to the analogue quantities in the LH-NMSSM scenario
in the vast part of the $(\lambda,\,\kappa)$-plane, (cf.\  
{Figure}~\ref{HIGLSPN1mass} for $m_{\tilde{\chi}^0_1}$ and
{Figure}~\ref{HIGLSPN1N2_350M9P55}{ for the 
corresponding
cross sections}).  

Regarding the Higgs sector, {it is
possible to get a MSSM counterpart with the same SM-Higgs mass and a
similar spectrum for the other Higgs states (with the exception of the
new singlet states) for
each point in the $(\lambda,\,\kappa)$-plane of the LH scenario.}

\begin{table}[h]
 \begin{center}
\begin{tabular}{|c|c|c|c|c|c|}\hline
$m_{\tilde{\chi}^0_1}$&$m_{\tilde{\chi}^0_2}$&$m_{\tilde{\chi}^0_3}$&$m_{\tilde{\chi}^0_4}$&$m_{\tilde{\chi}^{\pm}_1}$&$m_{\tilde{\chi}^{\pm}_2}$\\\hline
  114.8 GeV&123.3 GeV&454.4 GeV&1604.1 GeV&119.4 GeV&1604.1 GeV\\\hline
\end{tabular}             
\end{center}

\begin{center}
\begin{tabular}{|c||c|c|}\hline{\textbf{MSSM}}, $\sigma(e^+e^-\rightarrow\tilde{\chi}^0_1\tilde{\chi}^0_2)\,\,$  & $\sqrt{s}=350$~GeV & $\sqrt{s}=500$~GeV\\\hline\hline
$P=(-0.9,0.55)$&791.7 fb&391.4 fb\\\hline
$P=(0.9,-0.55)$&526.7 fb&261.7 fb\\\hline
\end{tabular}

\begin{tabular}{|c||c|c|}\hline{\textbf{MSSM}}, $\sigma(e^+e^-\rightarrow\tilde{\chi}^+_1\tilde{\chi}^-_1)$ & $\sqrt{s}=350$~GeV & $\sqrt{s}=500$~GeV\\\hline\hline
$P=(-0.9,0.55)$ &2348.8 fb & 1218.9 fb\\\hline
$P=(0.9,-0.55)$ &445.1 fb & 246.2 fb\\\hline
\end{tabular} \end{center}
 \caption{The reference MSSM scenario for {the} LH scenario:  neutralino and chargino masses [GeV] and production cross sections $\sigma(e^+e^-\rightarrow\tilde{\chi}^0_1\tilde{\chi}^0_2)$, $\sigma(e^+e^-\rightarrow\tilde{\chi}^+_1\tilde{\chi}^-_1)$  [fb].\label{HigLSPTab_MSSMXsecs}}
\end{table}

In Figure~\ref{HIGN1plots}, the NMSSM $\tilde{\chi}^0_1$ mass and its
singlino component are shown. A negligible singlino component
corresponds to a region in which the NMSSM $m_{\tilde{\chi}^0_1}$ is
very close to the MSSM value $m_{\tilde{\chi}^0_1}=114.8$ GeV. Vice
versa, with a higher singlino admixture the LSP mass,
$m_{\tilde{\chi}^0_1}$, within NMSSM significantly decreases.

Likewise, the neutralino polarised production cross sections
$\sigma(e^+e^-\rightarrow\tilde{\chi}^0_1\tilde{\chi}^0_2)$ decrease
with respect to the MSSM value following larger singlino component in
$\tilde{\chi}^0_1$, see Figure~\ref{HIGLSPN1N2_350M9P55}, as it is
expected since the singlino does not couple directly to the gauge
fields.  The tree-level NMSSM chargino masses and production
cross-sections,
$\sigma(e^+e^-\rightarrow\tilde{\chi}^+_1\tilde{\chi}^-_1)$, depend only
on $M_2$, $\mu_{\rm eff}$, $\tan\beta$, therefore chargino production
cross sections are identical to the MSSM values displayed in
Table~\ref{HigLSPTab_MSSMXsecs} along all the
$(\lambda,\,\kappa)$-plane.

\begin{figure}[htb]\centering
\includegraphics[width=.49\textwidth]{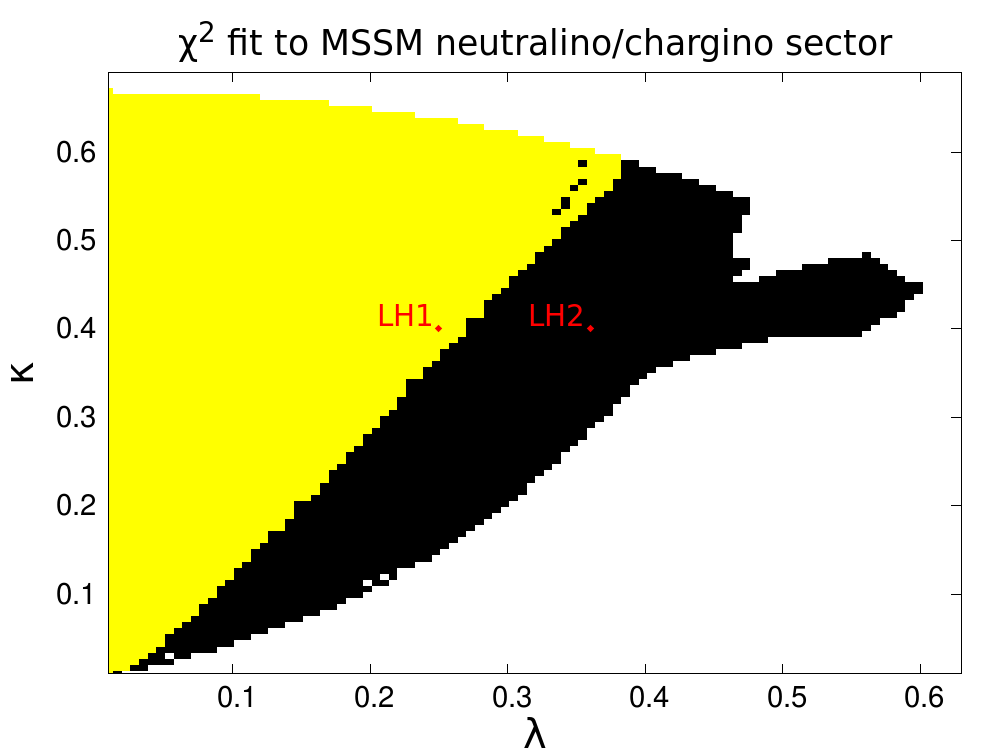}
\caption{LH scenario: fit to the MSSM. Yellow areas are compatible with the MSSM at 95\%~C.L., while black ones are excluded by the collider observables. The points LH1 $(\lambda,\kappa)=(0.25, 0.4)$ and LH2 $(\lambda,\kappa)=(0.36, 0.4)$ are also shown. \label{HigLSP_Fit}}
\end{figure}

According to the recipe in Section~\ref{Sec_Strategy}, we assume for
each point in the $(\lambda,\,\kappa)$-plane of the LH scenario the
experimental measurement of:
\begin{itemize}
 \item $m_{\tilde{\chi}^0_1}$, $m_{\tilde{\chi}^0_2}$ {and
       $m_{\tilde{\chi}^{\pm}_1}$} with  
an uncertainty of $0.5\%$.
 \item $\sigma(e^+e^-\rightarrow\tilde{\chi}^0_1\tilde{\chi}^0_2)$ for $P= (\mp0.9,\pm0.55)$ at $\sqrt{s} = 350$ and 500 GeV with $1\%$ uncertainty.
  \item $\sigma(e^+e^-\rightarrow\tilde{\chi}^+_1\tilde{\chi}^-_1)$, for $P= (\mp0.9,\pm0.55)$ at $\sqrt{s} = 350$ and 500 GeV with $1\%$ uncertainty.
\end{itemize}
In the regions in which the singlino component in $\tilde{\chi}^0_3$ is
higher, $m_{\tilde{\chi}^0_3}$ {is decreased} and
$\tilde{\chi}^0_2\tilde{\chi}^0_3$ may be kinematically accessible, see Figure~\ref{HIGLSPN2N3_500M9P55}. In
these cases, if $\tilde{\chi}^0_3$ is detectable through its decays, we
consider also $m_{\tilde{\chi}^0_3}$ and
$\sigma(e^+e^-\rightarrow\tilde{\chi}^0_2\tilde{\chi}^0_3)$. The
production $\tilde{\chi}^0_1\tilde{\chi}^0_3$ is negligible almost
everywhere.  With these assumptions, a $\chi^2$-fit to the MSSM gives
the result displayed in Figure~\ref{HigLSP_Fit}: the yellow areas
correspond to regions in the $(\lambda,\,\kappa)$-plane that are at
95\%~C.L. compatible with the MSSM, while in the black area MSSM is
excluded.
Therefore, a significant region of the parameter space, passing the
implemented phenomenological and experimental constraints,  can
definitely be distinguished from the MSSM using collider
observables. This is due to a higher 
singlino component in the neutralino $\tilde{\chi}^0_3$ (and partially in $\tilde{\chi}^0_1$ as well,
cf.\ Figure~\ref{HIGLSPN15}.

\begin{figure}[htb]\centering
\subfigure[]{\label{HigLSP_RedCoup}\includegraphics[width=.49\textwidth]{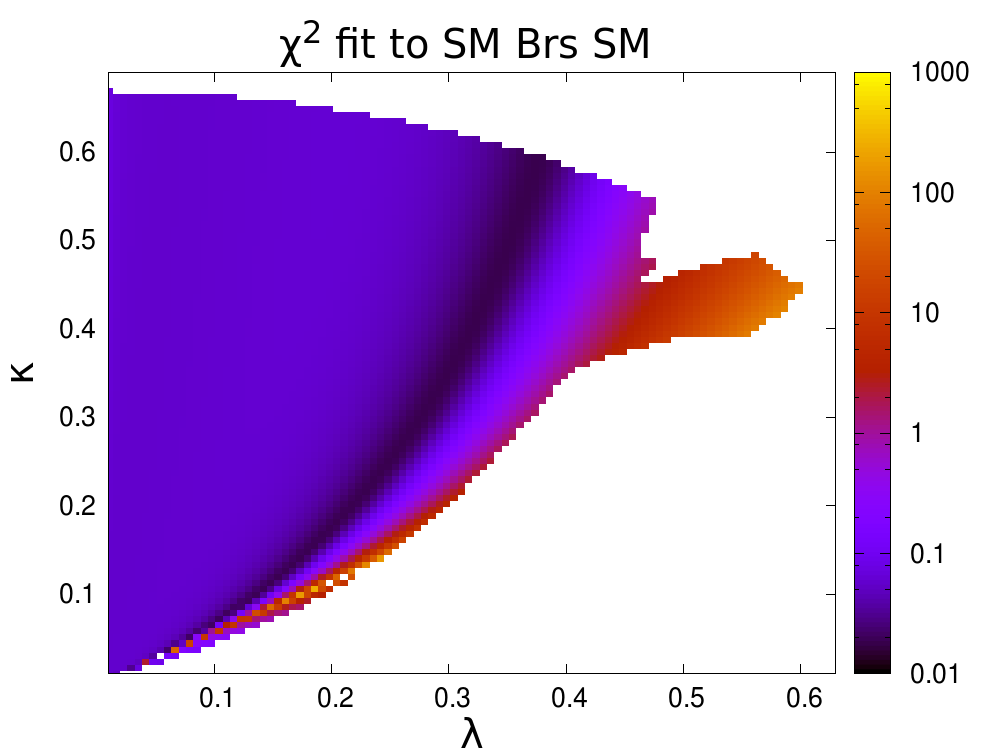}}
\subfigure[]{\label{HIG_SMHiggs}\includegraphics[width=.49\textwidth]{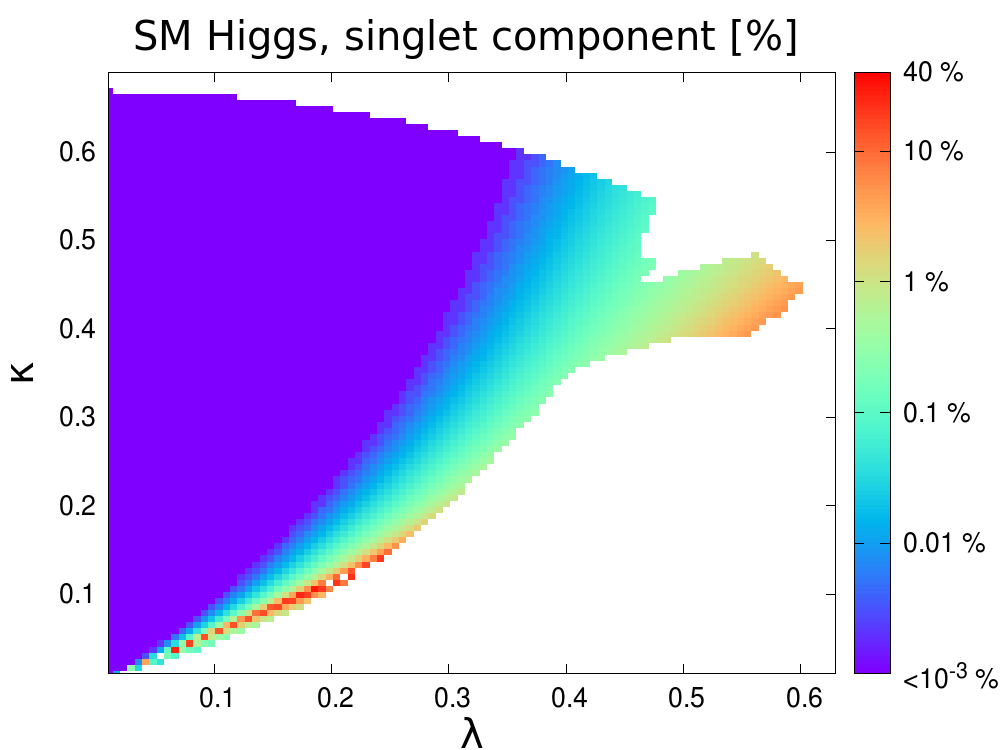}}
\caption{LH scenario: (a) 7-d.o.f. $\chi^2$-fit to the SM of the
 reduced couplings to $g,\gamma,W,Z,b,c,\tau$; (b) Singlet component in the SM-like Higgs, in \%. \label{HIG_SMHiggsfits}}
\end{figure}

We attempt here a reconstruction of the MSSM $M_1$, $M_2$, $\mu$, $\tan\beta$ and $m_{\tilde{\nu}_e}$ for two sample points in the $(\lambda,\kappa)$-plane of the LH scenario, relatively close to the boundary between the regions of compatibility from Figure~\ref{HigLSP_Fit}.

\begin{itemize}
 \item The point LH1,  with $(\lambda,\kappa)=(0.25,\, 0.4)$, features the
        masses and cross sections given in Tables~\ref{LH1masses} and~\ref{LH1xsecs}.
The fit to LH1 turns out to be compatible with the MSSM, $\chi^2=1.1$, and yields
\begin{align}
 &M_1=360\pm 40 \mbox{ GeV,\,\,\, }\,\,M_2=1300 \pm 300\mbox{ GeV, }\nonumber\\
 &\mu_{\rm eff}=124\pm 2\mbox{ GeV, }\,\,\,\tan\beta \leq 4,\nonumber \\
 &m_{\tilde{\nu}_e}\leq 470\; \mathrm{GeV}\,.
\end{align}

\item For our second example, the point LH2  with
	 $(\lambda,\kappa)=(0.36,\, 0.4)$ is taken and the corresponding masses and
	 cross sections are given in Tables~\ref{LH2masses} and~\ref{LH2xsecs}. The point LH2 in not compatible with the MSSM, with the fit giving $\chi^2=1700$ and the following parameter values:
\begin{align}
 &M_1\;\; \mathrm{unconstrained,}\qquad\quad\;\,\,\, M_2=317.0\pm0.5\mbox{ GeV, }\nonumber\\
 &\mu_{\rm eff}=129.3\pm0.6\mbox{ GeV, }\quad \,\,\,\tan\beta < 1.1, \nonumber\\
 & m_{\tilde{\nu}_{e}}=297\pm 15\; \mathrm{GeV}.
\end{align}

 \end{itemize}

\begin{table}
 \begin{center}
\begin{tabular}{|c|c|c|c|c|c|c|}\hline
$m_{\tilde{\chi}^0_1}$&$m_{\tilde{\chi}^0_2}$&$m_{\tilde{\chi}^0_3}$&$m_{\tilde{\chi}^0_4}$&$m_{\tilde{\chi}^0_5}$&$m_{\tilde{\chi}^{\pm}_1}$&$m_{\tilde{\chi}^{\pm}_2}$\\\hline
 111.6 GeV& 125.2 GeV   &389.0 GeV &454.4 GeV& 1604 GeV & 119.4 GeV & 1604 GeV\\\hline
\end{tabular}             \end{center}  
\caption{Neutralino and chargino masses {in the
 light higgsino scenario for the
 reference point LH1 with $(\lambda,\kappa)=$(0.25, 0.4)}.\label{LH1masses}}
\end{table}
\normalsize 
\begin{table}
\begin{center}
\begin{tabular}{|c||c|c|}\hline$\sigma(e^+e^-\rightarrow\tilde{\chi}^0_1\tilde{\chi}^0_2)$  & $\sqrt{s}=350$~GeV & $\sqrt{s}=500$~GeV\\\hline\hline
$P=(-0.9,0.55)$&  781.5 fb& 385.8 fb\\\hline
$P=(0.9,-0.55)$&519.9  fb& 257.9 fb\\\hline
\end{tabular} \end{center}
\caption{Neutralino production cross sections {in
 the light higgsino scenario, reference point LH1 with $(\lambda,\kappa)=$(0.25, 0.4)}.\label{LH1xsecs}}
\end{table}

\begin{table}
 \begin{center}
\begin{tabular}{|c|c|c|c|c|c|c|}\hline
$m_{\tilde{\chi}^0_1}$&$m_{\tilde{\chi}^0_2}$&$m_{\tilde{\chi}^0_3}$&$m_{\tilde{\chi}^0_4}$&$m_{\tilde{\chi}^0_5}$&$m_{\tilde{\chi}^{\pm}_1}$&$m_{\tilde{\chi}^{\pm}_2}$\\\hline
 104.2 GeV& 128.4 GeV   &282.4 GeV &454.4 GeV&1604 GeV & 119.4 GeV & 1604 GeV\\\hline
\end{tabular}             \end{center} %
\caption{Neutralino and chargino masses {in the
 light higgsino scenario for the
 reference point LH2 with $(\lambda,\kappa)=$(0.36, 0.4)}.\label{LH2masses}}
 \end{table}
\normalsize
\begin{table}
\begin{center}
\begin{tabular}{|c||c|c|}\hline$\sigma(e^+e^-\rightarrow\tilde{\chi}^0_1\tilde{\chi}^0_2)$  & $\sqrt{s}=350$~GeV & $\sqrt{s}=500$~GeV\\\hline\hline
$P=(-0.9,0.55)$&739.0 fb&363.3 fb\\\hline
$P=(0.9,-0.55)$&491.5 fb&242.8 fb\\\hline
\end{tabular} 

\begin{tabular}{|c||c|c|}\hline$\sigma(e^+e^-\rightarrow\tilde{\chi}^0_2\tilde{\chi}^0_3)$  & $\sqrt{s}=350$~GeV & $\sqrt{s}=500$~GeV\\\hline\hline
$P=(-0.9,0.55)$&  not accessible &15.4 fb\\\hline
$P=(0.9,-0.55)$& not accessible &10.4 fb\\\hline
\end{tabular} \end{center}
\caption{Neutralino production cross sections {in the
 light higgsino scenario for the
 reference point LH2 with $(\lambda,\kappa)=$(0.36, 0.4)}.\label{LH2xsecs} }
\end{table}
 
Additional information from the heavier neutralino states, such as
$\tilde{\chi}^0_3$ (if its production is not already kinematically
allowed at 500 GeV) or $\tilde{\chi}^0_4$, may help in reducing the region
compatible with the MSSM. For example, given a $(\lambda,\kappa)$
coordinate and the corresponding $M_1$, $M_2$, $\mu$, $\tan\beta$
reconstructed from the fit, one can derive the masses of the heavier
states and look for them at higher energies at the ILC or at the LHC,
either confirming the fit to the MSSM or pinpointing the NMSSM.

As suggested in Section~\ref{Sec_Strategy}, our study may be extended by including information from the Higgs sector. A result of the na\"{\i}ve $\chi^2$-fit
to the SM of the Higgs reduced couplings  
to $g,\gamma,W,Z,b,c,\tau$, each defined as a ratio $g_h/g_{h_{\rm SM}}$ between the SM-like Higgs coupling to the corresponding SM Higgs coupling,
is shown in Figure~\ref{HigLSP_RedCoup}.\footnote{We used the expected
accuracies for the SM-like Higgs boson branching ratios $\Delta
\mbox{Br}/ \mbox{Br}$ from \cite{Baer:2013cma}. }  In large part of the
$(\lambda,\kappa)$-plane, the SM-like Higgs of the LH scenario is
compatible with the SM ($\chi^2\lesssim 14$), corresponding to the
MSSM-like area from the fit in Figure~\ref{HigLSP_Fit}. 
A SM-like Higgs with a higher singlet component, see Figure~\ref{HIG_SMHiggs}, 
corresponds to a worse fit: there are two regions that are not compatible with
the SM and have a different behaviour with respect of MSSM-like
areas. The conclusion from this fit is therefore consistent with that of
{Figure}~\ref{HigLSP_Fit} without clearly improving our analysis.

\begin{figure}[htb]\centering
\subfigure[]{\label{HigLSP_Neu3toA1}\includegraphics[width=.49\textwidth]{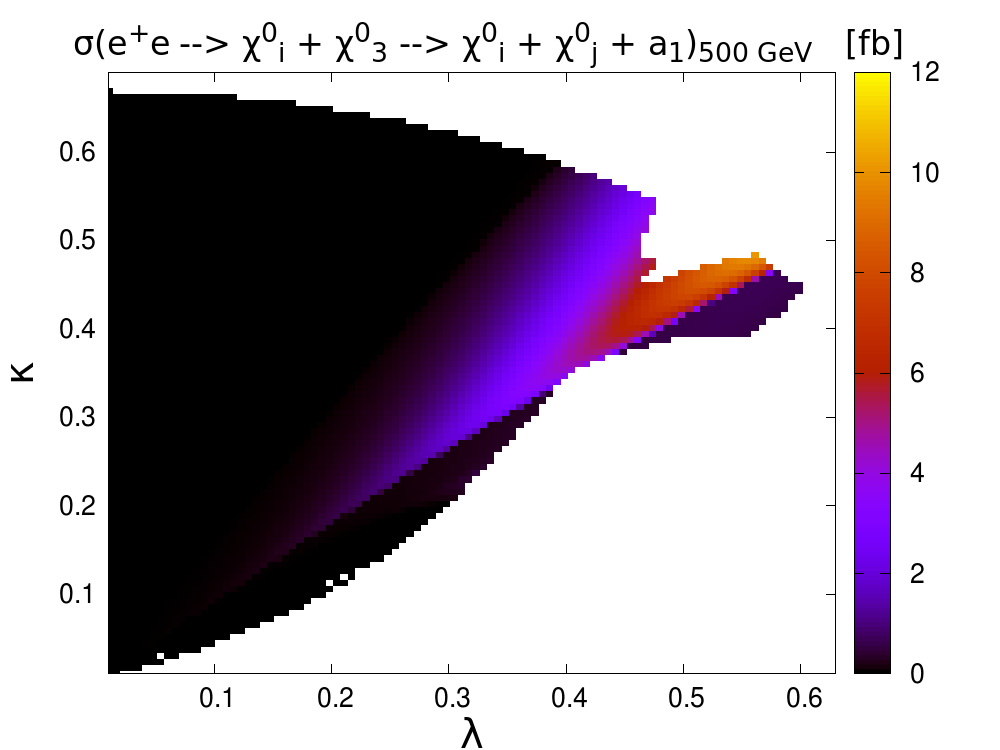}}
\subfigure[]{\label{HIGLSP_CPodd1mass}\includegraphics[width=.49\textwidth]{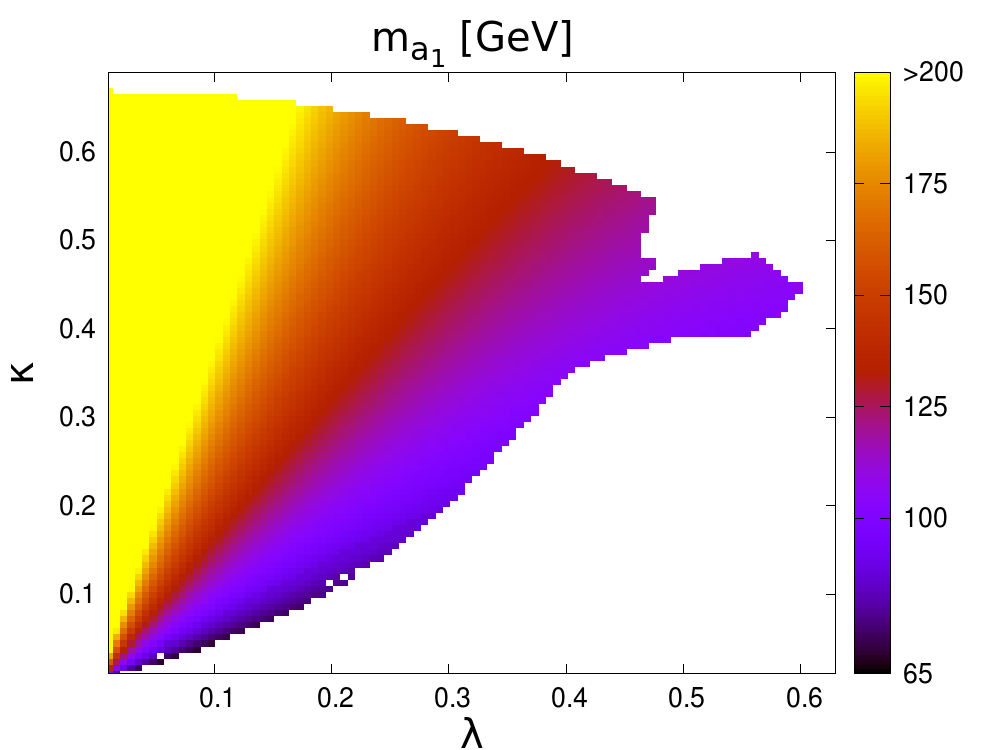}}
\caption{LH scenario: (a) inclusive cross section $e^+e^-\rightarrow\tilde{\chi}^0_i\tilde{\chi}^0_3\rightarrow\tilde{\chi}^0_i\tilde{\chi}^0_ja_1$ [fb], with $i,\,j=1,2$; (b) lightest CP-odd Higgs mass $m_{a_1}$ [GeV].\label{HIGA1_prod_plots}}
\end{figure}

Additional information about the NMSSM Higgs sector could obtained if new singlets are directly visible. This possibility opens up in a region with a higher singlino component in $\tilde{\chi}^0_3$, where the decays $\tilde{\chi}^0_3\rightarrow\tilde{\chi}^0_{1,2}a_1$ become open. If the production cross section for $\tilde{\chi}^0_3$ is non-negligible one could observe the pseudoscalar $a_1$ via its decays $a_1 \to b \bar{b}$. In Figure~\ref{HigLSP_Neu3toA1} we show an inclusive cross section for production of $a_1$, where both production modes, $e^+ e^- \to \tilde{\chi}^0_1 \tilde{\chi}^0_3$ and $e^+ e^- \to \tilde{\chi}^0_2 \tilde{\chi}^0_3$, for both polarisations has been added up together with decays $ \tilde{\chi}^0_3 \to \tilde{\chi}^0_1 a_1 $ and $ \tilde{\chi}^0_3 \to \tilde{\chi}^0_2 a_1 $. In certain regions of parameter space, with cross sections of order 10~fb, the new state should be clearly visible. This could serve as confirmation of the NMSSM, since the MSSM would be 
already excluded by the fit to other observables. As a reference, in Figure~\ref{HIGLSP_CPodd1mass} we also show the mass of the pseudoscalar $a_1$.

\subsection{Light gaugino scenario, $\mu_{\rm eff}>M_1>M_2$ \label{SubsecGAULSP}}

Finally, we study an NMSSM scenario with light gauginos (LG), whose
neutralino/chargino sector is given by:
\begin{equation}
 M_1=240\mbox{ GeV,\,\,\, }\,\,\,M_2=105\mbox{ GeV, }\,\,\,\mu=\mu_{\rm eff}=505\mbox{ GeV, }\,\,\,	\tan\beta=9.2\,,\label{GauLSP_parameters}
\end{equation}
with $\lambda\in[0,0.7]$ and $\kappa\in[0,0.7]$. The singlet soft
 trilinear parameters are $A_{\lambda}=3700$ GeV,
 $A_{\kappa}=-40$~GeV. The first generation sfermion masses
 are\begin{equation}\label{GauLSP_sfMass} m_{\tilde{e}_L}=303.4\mbox{
 GeV},\,\,\,
 m_{\tilde{e}_L}=m_{\tilde{e}_R},\,\,\,m_{\tilde{\nu}_e}^2=m_{\tilde{e}_L}^2+\cos(2\beta)\cos^2\theta_Wm_Z^2,
\end{equation} while squarks masses are $>1$ TeV. For the full set of soft parameters, we refer to Appendix~\ref{Appendix:lightGaugino}.

\begin{figure}[t!]
\begin{center}
 \includegraphics[width=3.5in]{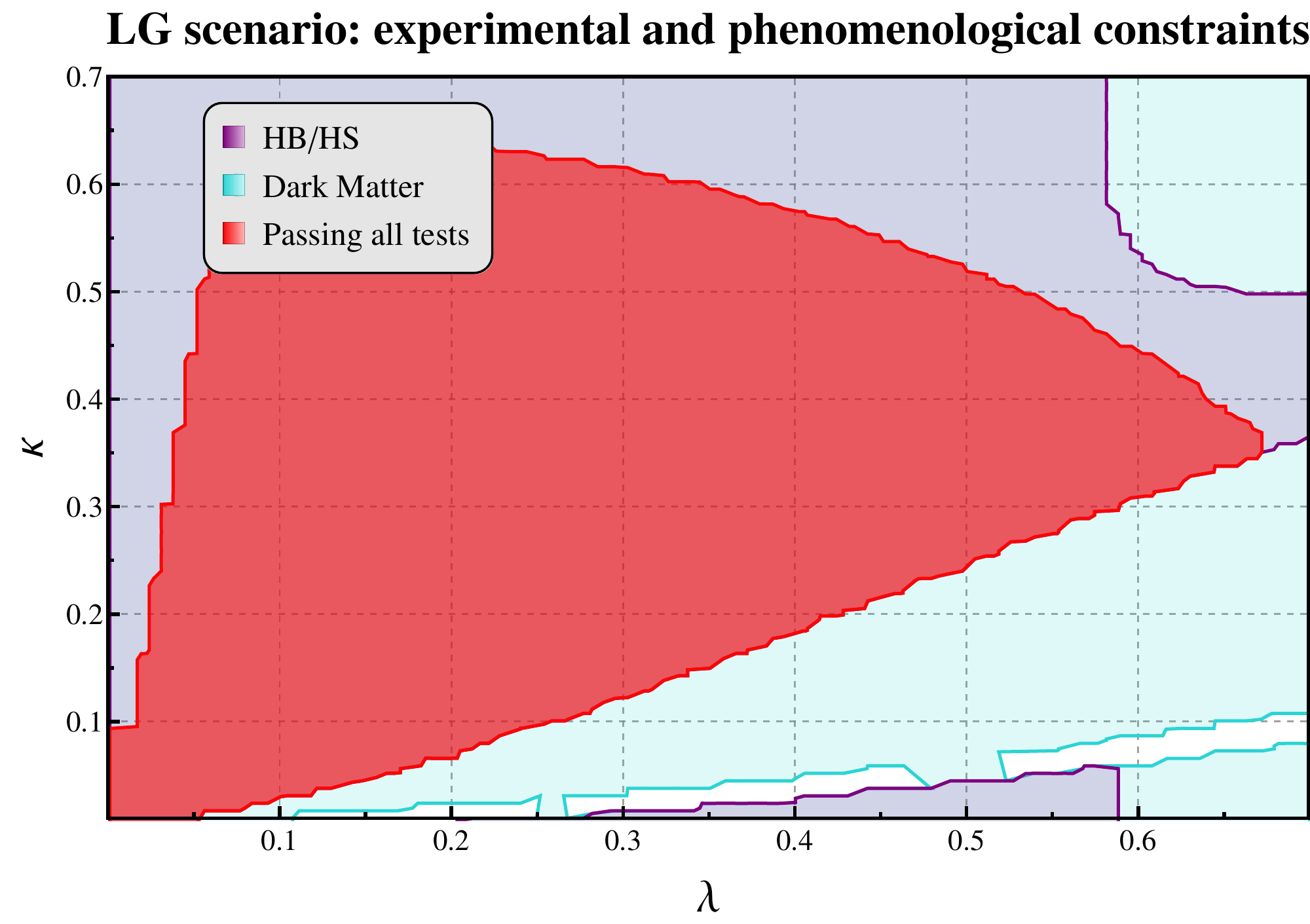}
 \caption{The light gaugino scenario: regions in the $(\lambda,\kappa)$-plane allowed by experimental and phenomenological constraints. The light-blue region  passes the dark matter constraints. The purple-coloured region passes checks from \texttt{HiggsBounds} and \texttt{HiggsSignals}. The areas allowed by all the constraints are shown in red.\label{GauLSP_Exclusions}}
  \end{center}
\end{figure}

In Figure~\ref{GauLSP_Exclusions} we display the result of the scan in
the NMSSM $(\lambda,\kappa)$-plane after our tests implemented within
\texttt{NMSSMTools}, \texttt{HiggsBounds} and \texttt{HiggsSignals}. The
colour conventions are the same as for the LH scenario,
Section~\ref{SubsecHIGLSP}; for the LG scenario the regions allowed by
the Higgs sector constraints from \texttt{HiggsBounds} and
\texttt{HiggsSignals} overlap entirely those passing DM matter
constraints.

A reference MSSM scenario with an almost indistinguishable lighter
(tree-level) neutralino and chargino mass spectrum and production cross
sections is found by choosing $M_1$, $M_2$, $\mu$, $\tan\beta$ and the
first generation slepton masses as in Eq.~\eqref{GauLSP_parameters}, see
Table~\ref{GauLSPTab_MSSMXsecs}.

\begin{table}[htb]
 \begin{center}
\begin{tabular}{|c|c|c|c|c|c|}\hline
$m_{\tilde{\chi}^0_1}$&$m_{\tilde{\chi}^0_2}$&$m_{\tilde{\chi}^0_3}$&$m_{\tilde{\chi}^0_4}$&$m_{\tilde{\chi}^{\pm}_1}$&$m_{\tilde{\chi}^{\pm}_2}$\\\hline
 99.5 GeV& 237.0 GeV   &510.1 GeV   &518.7 GeV  & 99.6 GeV & 518.7 GeV\\\hline
\end{tabular}             
\end{center}

\normalsize
\begin{center}
\begin{tabular}{|c||c|c|}\hline{\textbf{MSSM}}, $\sigma(e^+e^-\rightarrow\tilde{\chi}^0_1\tilde{\chi}^0_2)$  &$\sqrt{s}=350$~GeV & $\sqrt{s}=500$~GeV\\\hline\hline
$P=(-0.9,0.55)$&7.3 fb&113.4 fb\\\hline
$P=(0.9,-0.55)$&0.1 fb&1.8 fb\\\hline
\end{tabular} \end{center}\normalsize
\begin{center}
\begin{tabular}{|c||c|c|}\hline{\textbf{MSSM}}, $\sigma(e^+e^-\rightarrow\tilde{\chi}^+_1\tilde{\chi}^-_1)$  &$\sqrt{s}=350$~GeV & $\sqrt{s}=500$~GeV\\\hline\hline
$P=(-0.9,0.55)$&2692.1 fb&1252.6 fb\\\hline
$P=(0.9,-0.55)$&44.5 fb&19.4 fb\\\hline
\end{tabular} \end{center}
 \caption{The reference light gaugino  MSSM scenario:
neutralino and chargino masses [GeV] and production cross sections $\sigma(e^+e^-\rightarrow\tilde{\chi}^0_1\tilde{\chi}^0_2)$, $\sigma(e^+e^-\rightarrow\tilde{\chi}^+_1\tilde{\chi}^-_1)$  [fb].\label{GauLSPTab_MSSMXsecs}}
\end{table}

\begin{figure}[htb]\centering
\includegraphics[width=.49\textwidth]{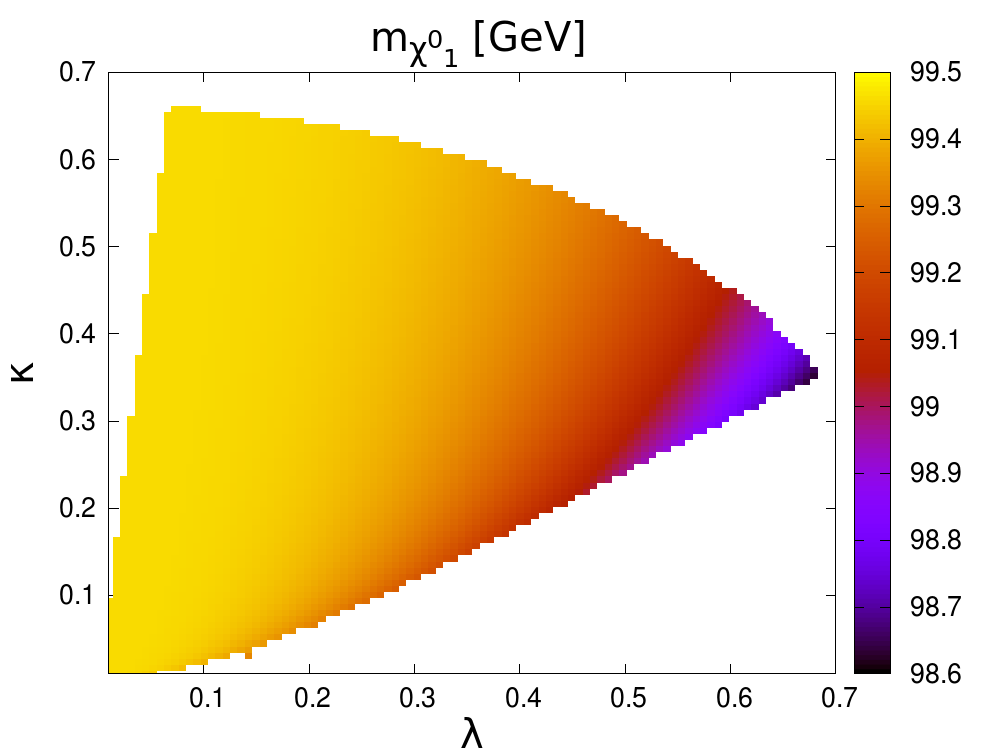}
\caption{The LG scenario: the mass $m_{\tilde{\chi}^0_1}$ [GeV].\label{GAULSPN1mass}}
\end{figure}

In the LG scenario $m_{\tilde{\chi}^0_1}$ is very close to the reference
MSSM value 99.5~GeV and it varies very mildly in the
$(\lambda,\,\kappa)$-plane as the singlino component in
$\tilde{\chi}^0_1$ is approximately zero, see
Figure~\ref{GAULSPN1mass}. A similar reasoning applies to the production
cross section
$\sigma(e^+e^-\rightarrow\tilde{\chi}^0_1\tilde{\chi}^0_2)$, while the
chargino production,
$\sigma(e^+e^-\rightarrow\tilde{\chi}^+_1\tilde{\chi}^-_1)$, is exactly
identical at the tree-level as explained in Section~\ref{SubsecHIGLSP}.

\begin{figure}[htb]\centering
\subfigure[]{\label{GauLSP_Fit}\includegraphics[width=.49\textwidth]{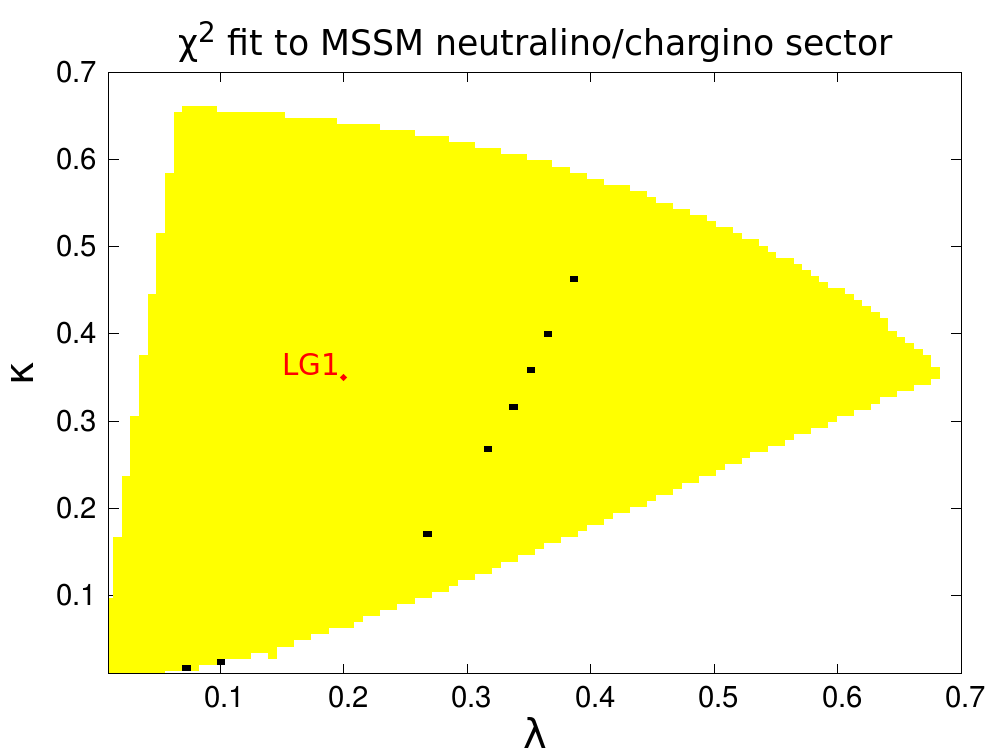}}
\subfigure[]{\label{GauLSP_RedCoup}\includegraphics[width=.49\textwidth]{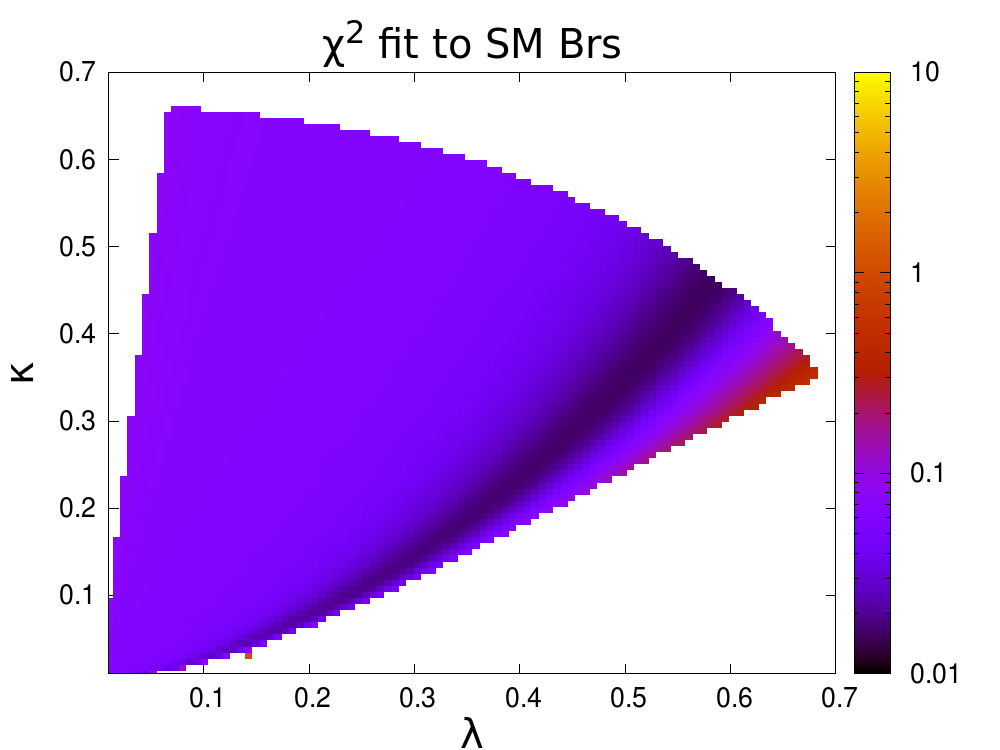}}
\caption{LG scenario: (a) fit to the MSSM. Yellow areas are compatible with the MSSM at 95\%~C.L., while black ones are excluded by the collider observables. The point LG1 $(\lambda,\kappa)=$(0.2, 0.35) is displayed. (b) $\chi^2$-fit to the SM of the reduced Higgs boson couplings to $g,\gamma,W,Z,b,c,\tau$. \label{Gau_fits}}
\end{figure}

We only use cross sections larger than 1 fb for $\chi^2$-fit to the
MSSM. Figure~\ref{GauLSP_Fit} shows that our fit alone is not able
to distinguish in this case between the two models, as basically every
point in the allowed region is compatible with the MSSM.

As example we analyse the point LG1 with $(\lambda,\kappa)=(0.2, 0.35)$ and
the remaining parameters given by Eqs.~\eqref{GauLSP_parameters}
and~\eqref{GauLSP_sfMass} that features the masses and cross
sections listed in Tables~\ref{LG1masses} and~\ref{LG1xsecs}. For $P=(0.9,-0.55)$ the cross section
  $\sigma(e^+e^-\rightarrow\tilde{\chi}^0_1\tilde{\chi}^0_2)$ 
 at $\sqrt{s}=350$~GeV {is below 1 fb} and {the process}
 $e^+e^-\rightarrow\tilde{\chi}^0_1\tilde{\chi}^0_3$ is kinematically
 not allowed {for} both at 350 and 500 GeV. 
The {remaining observables lead to a fit that} 
is compatible with the
 MSSM giving $\chi^2$=0.07:
\begin{align}
 &M_1=239.9\pm0.9\mbox{ GeV,\,\,\, }\,\,M_2=104.4\pm0.8\mbox{ GeV, }\nonumber \\
 &\mu_{\rm eff}=504.7\pm47.6\mbox{ GeV, }\,\,\,\tan\beta=11.4\pm2.8,\nonumber \\
 & m_{\tilde{\nu}_e}=292.8\pm3.9\; \mathrm{GeV}\,.
\end{align}
These values are remarkably close to {the `true'} 
input parameters given by Eqs.~\eqref{GauLSP_parameters} 
and~\eqref{GauLSP_sfMass}. A na\"{\i}ve fit of the SM-like Higgs reduced couplings does not provide
information useful for model distinction, see
Figure~\ref{GauLSP_RedCoup}, as they are always compatible with the SM,
unlike in the LH scenario.

 \begin{table}[htb]
 \begin{center}
\begin{tabular}{|c|c|c|c|c|c|c|}\hline
$m_{\tilde{\chi}^0_1}$&$m_{\tilde{\chi}^0_2}$&$m_{\tilde{\chi}^0_3}$&$m_{\tilde{\chi}^0_4}$&$m_{\tilde{\chi}^0_5}$&$m_{\tilde{\chi}^{\pm}_1}$&$m_{\tilde{\chi}^{\pm}_2}$\\\hline
 99.4 GeV& 237.0 GeV   &510.4 GeV &518.3 GeV&1768.2 GeV & 99.5 GeV & 518.7 GeV\\\hline
\end{tabular}             \end{center} 
\caption{Neutralino and chargino masses {in the
  light gaugino scenario for the reference point LG1 with $(\lambda,\kappa)=$(0.2, 0.35)}.\label{LG1masses}}
 \end{table}

\normalsize

\begin{table}[htb]
\begin{center}
\begin{tabular}{|c||c|c|}\hline$\sigma(e^+e^-\rightarrow\tilde{\chi}^0_1\tilde{\chi}^0_2)$  & $\sqrt{s}=350$~GeV & $\sqrt{s}=500$~GeV\\\hline\hline
$P=(-0.9,0.55)$&7.3 fb&113.5 fb\\\hline
$P=(0.9,-0.55)$&0.1 fb&1.8 fb\\\hline
\end{tabular} \end{center}\normalsize
\caption{Neutralino production cross sections {in the
  light gaugino scenario for the reference point LG1 with $(\lambda,\kappa)=$(0.2, 0.35)}.\label{LG1xsecs} }
\end{table}

This behaviour can be understood by analysing the mixing within the
neutralino sector. In the NMSSM, the singlino does not mix directly with
gauginos but only indirectly via higgsino states, see
Eq.~\eqref{NMSSMneutralinoMassMatrix} and
Appendix~\ref{App:CNsectors}. If, like in the LG scenario, $\mu_{\rm eff}
\gg M_1, M_2$, the mixing remains small even for a relatively light
singlino. Therefore the properties of the light chargino and 
neutralino
states, including masses and cross sections, remain very similar
throughout the $(\lambda,\kappa)$-plane and cannot be distinguished from
the MSSM case. In contrast to that, in the light singlino scenario from
Section~\ref{Subsection:LightSinglino}, $M_1 = 365$~GeV and $\mu_{\rm
eff} = 484$~GeV are of the similar size resulting in significant mixing:
$\tilde{\chi}^0_2 \simeq 22\%\, \tilde{B} + 73\%\, \tilde{S}$ and
$\tilde{\chi}^0_3\simeq 72\%\, \tilde{B} + 25\%\, \tilde{S}$. Since in
the LS case the singlino component
makes up a significant part of the light neutralinos,
the modification of the couplings allows the clear discrimination from the
MSSM.

\section{Conclusions and outlook}\label{Sec_Conclusions}
It will be very important to develop methods
how to discriminate between the NMSSM and the MSSM at
future experiments, as the two models may reproduce experimentally very similar
light Higgs sectors as well as lower supersymmetric spectra. In this
paper we have outlined a model distinction strategy that focuses on the
neutralino and chargino sector and we have applied it to a series of
NMSSM scenarios with different singlino, gaugino and higgsino properties.
The idea is {to assume} that the 
lightest neutralino and
chargino masses 
as well as their polarised pair
production cross sections are measurable 
at a linear collider and to reconstruct the
corresponding MSSM parameters, $M_1,\,M_2,\,\mu,\tan\beta$,
{via a}
$\chi^2$-fit. {In case such a fit} 
clearly excludes the MSSM  hypothesis it would strongly point 
towards an extended model, preferably the NMSSM. Integrating the
analysis with further information from the Higgs sector or from heavier
neutralino resonances could confirm such a new model hypothesis.  
Throughout our study we have assumed to
operate at the ILC with two different energy stages, namely
at $\sqrt{s}=350$ and 500 GeV, using electron and
positron beam polarisation with $P=(\pm0.9,\mp0.55)$.

We have introduced three classes of scenarios with different phenomenological
aspects concerning the model distinction: a light singlino, a light higgsino
and a light gaugino scenario.
We have {first} analysed {an NMSSM scenario} with
singlino components in the $\tilde{\chi}^0_2$ and a  wino-like LSP
$\tilde{\chi}^0_1$, i.e.\ with an inverted hierarchy of the gaugino
mass parameters.
Such a NMSSM scenario does not result in set of observables consistent with the
MSSM. Accessing the mixing character of
the heavier neutralino $\tilde{\chi}^0_3$ would confirm the situation
and point to a model with an extended neutralino sector with respect to
the MSSM.

In the class with light
higgsinos, one usually has the hierarchy $\mu_{\rm eff}<M_1<M_2$. In the corresponding NMSSM parameter
space, a large part of the
$(\lambda,\kappa)$-plane features the heavy and decoupled singlino while the
$\tilde{\chi}^0_1$
is higgsino-like. Such a model is indistinguishable from the
MSSM. However, if a sufficient singlino admixture is present in the light neutralinos, the neutralino sector changes appreciably, allowing for a discrimination between the MSSM and the NMSSM. In some region of the parameter space additional pseudoscalar Higgs $a_1$ could also be observed. Precise measurement of the SM-like Higgs
couplings would be beneficial  for
a confirmation of these conclusions.

As a third
class we have chosen light gaugino scenarios again with an inverted
hierarchy
$M_2<M_1<\mu_{\rm eff}$ but with $\mu_{\rm eff}-M_1\sim\mathcal{O}(250)$
GeV. In this way the singlino does not 
significantly mix with gauginos in the lightest neutralino states. 
In the light of our experimental
assumptions, the low mass spectrum and production cross sections are not
distinguishable from the MSSM ones all over the allowed
$(\lambda,\kappa)$-plane. In this case analysing the
SM-like Higgs couplings also does not provide
further information.

Our studies show that 
the neutralino and chargino sector can provide the crucial information for the model distinction between the MSSM and
the NMSSM. Such a
discrimination depends on the gaugino mass
hierarchies and the actual singlino admixture in the light neutralino states. Precise measurements and a model-independent
analysis for the determination of  the fundamental SUSY parameters are essential.

\section*{Acknowledgements}

The authors are thankful to M.~Berggren, F.~Domingo, P.~Gunnellini, J.~List,
O.~St\aa{}l, M.~Tonini, M.~de Vries, G.~Weiglein and L.~Zeune for useful
discussions.  S.~P. has been supported by DFG through the grant SFB 676
``Particles, Strings, and the Early Universe''. This work has been
partially supported by the MICINN, Spain, under contract FPA2010-17747;
Consolider-Ingenio CPAN~CSD2007-00042.  We thank also the Comunidad de
Madrid through Proyecto HEPHACOS S2009/ESP-1473 and the European
Commission under contract PITN-GA-2009-237920.

\appendix

\section{Chargino, neutralino and Higgs sector\label{App:Sectors}}

\subsection{Chargino and neutralino mass matrices}\label{App:CNsectors}

The tree-level chargino sector is identical for the MSSM and NMSSM. In
the $(\tilde{W}^{\pm}, \tilde{H}^{\pm})$ basis, the chargino mass matrix
reads 

\begin{equation}
 \mathcal{M}_C=\left(\begin{array}{cc}
      M_2&\sqrt{2}\,m_Z\cos\theta_W\cos \beta\\
      \sqrt{2}\,m_Z\cos\theta_W\sin \beta& \mu
      \end{array}\right),
\end{equation}
in the convention according to which $\tilde{\chi}^{-}$ is taken as the
particle and $\tilde{\chi}^{+}$ as its antiparticle (i.e. different
convention as in e.g. \cite{Haber:1984rc}). $M_2$ is chosen real and positive.
The charginos, eigenstates of $\mathcal{M}_C$, can be written as \cite{Desch:2003vw}

\begin{equation}\label{CharMix}
\left(\begin{array}{c}
      \tilde{\chi}^{-}_1\\
       \tilde{\chi}^{-}_2
      \end{array}\right)_{L,R}= U_{L,R}\left(\begin{array}{c}
\tilde{W}^{-}\\
       \tilde{H}^{-}
      \end{array}\right)_{L,R}=
      \left(\begin{array}{cc}
      \cos\Phi_{L,R}&\sin\Phi_{L,R}\\
      -\sin\Phi_{L,R}&\cos\Phi_{L,R}
      \end{array}\right)\left(\begin{array}{c}
\tilde{W}^{-}\\
       \tilde{H}^{-}
      \end{array}\right)_{L,R} \,,
\end{equation}
such that \begin{eqnarray}
         &&  m^2_{\tilde{\chi}^{\pm}_{1,2}} =\frac{1}{2}(M_2^2+\mu^2+2m_W^2\mp\Delta_C) \,,\\
  && \cos2\Phi_{L,R}=-(M_2^2-\mu^2\mp2m_W^2\cos2\beta)/\Delta_C \label{Cos2Phi}\,,
  \end{eqnarray}
where $\Delta_C=[(M_2^2-\mu^2)^2+4m_W^4\cos^22\beta+4m_W^2(M_2^2+\mu^2)+8m_W^2M_2\mu\sin2\beta]^{1/2}$.

\

The tree-level MSSM neutralino mass matrix in the $(\tilde{B}, \tilde{W}^0, \tilde{H}_d, \tilde{H}_u)$ basis,
\begin{equation}
 \mathcal{M}_{\rm MSSM}=\left(\begin{array}{cccc}
      M_1&0&-\cos\beta\sin\theta_Wm_Z&\sin\beta\sin\theta_Wm_Z\\
      0&M_2&\cos\beta\cos\theta_Wm_Z&-\sin\beta\cos\theta_Wm_Z\\
      -\cos\beta\sin\theta_Wm_Z&\cos\beta\cos\theta_Wm_Z&0&-\mu\\
      \sin\beta\sin\theta_Wm_Z&-\sin\beta\cos\theta_Wm_Z&-\mu&0
      \end{array}\right),
\end{equation}
can be diagonalised by a unitary matrix $N$, obtaining the neutralino eigenvectors and their masses:
\begin{equation}
 N^{\ast}\mathcal{M}_{\rm MSSM}N^{\dagger}=\mbox{diag}\{m_{\tilde{\chi}^0_1},\dots,m_{\tilde{\chi}^0_4}\}\mbox{.}
\end{equation}
 $\mathcal{M}_{\rm MSSM}$ is equivalent to the upper left block of the  the tree-level ($\mathbb{Z}_3$-invariant) NMSSM neutralino mass matrix, in the basis $(\tilde{\gamma},\tilde{Z}, \tilde{H}_d, \tilde{H}_u, \tilde{S})$ \cite{Ellwanger:2009dp}:  

\newcommand*{\tempbA}{\multicolumn{1}{|c}{{0}}}
\newcommand*{\tempbB}{\multicolumn{1}{|c}{{-\lambda v\sin\beta}}}
\newcommand*{\tempbC}{\multicolumn{1}{|c}{{-\lambda v\cos\beta}}}

\begin{equation}
 \Large\mathcal{M}_{\rm NMSSM}\normalsize=\left(\begin{array}{ccccc}
	\multicolumn{4}{c}{\multirow{4}{*}{\Large$\mathcal{M}_{\rm MSSM}$\normalsize}}&\tempbA\\
      &&&&\tempbA\\
      &&&&\tempbB\\
      &&&&\tempbC\\\cline{1-4}
      {0}&{0}&{-\lambda v\sin\beta}&{-\lambda v\cos\beta}&{-2\kappa\, \mu_{\rm eff}/\,\lambda}
      \end{array}\right)\,,\label{NMSSMneutralinoMassMatrix}
\end{equation}
with the only difference that now $\mu$ is substituted by $\mu_{\footnotesize\mbox{eff}\normalsize}=\lambda s$, where $s$ the vev of the singlet, and where $v_u^2+v_d^2=v^2=2m_Z^2/(g_1^2+g_2^2)\approx(174\mbox{ GeV})^2$. The NMSSM neutralino sector depends on two more singlet/singlino parameters with respect to the MSSM: $\lambda,\kappa$, while $\mu$ dependence is substituted by the dependence on the singlet vev $s$.

\subsection{$\mathbb{Z}_3$-NMSSM Higgs sector}
According to Ref.~\cite{Ellwanger:2009dp}, for the $\mathbb{Z}_3$-invariant NMSSM, the part of the superpotential describing Higgs-Singlet (self) interactions is given by:
\begin{equation}
 W_{\footnotesize\mbox{Higgs-singlet}\normalsize}=\lambda \hat{S}\hat{H}_u\cdot\hat{H}_d+\frac{\kappa}{3}\hat{S}^3\,,\label{Superpotential:HiggsSinglet}
\end{equation}
while the Yukawa couplings are described by
\begin{equation}
 W_{\footnotesize\mbox{Yukawa}\normalsize}=h_u \hat{Q}\cdot \hat{H}_u\hat{U}^c_R +h_d\hat{H}_d\cdot\hat{Q}\hat{D}^c_R+h_e \hat{H}_d\cdot\hat{L}\,\hat{E}^c_R\,\,.
\end{equation}
The Higgs soft SUSY breaking lagrangian reads:
\begin{align}
 -\mathcal{L}_{\footnotesize\mbox{Higgs-Singlet soft}\normalsize}=
                                                    h_u A_u Q\cdot H_u U^c_R-h_dA_dQ\cdot D^c_R-h_e A_e L\cdot H_d E^c_R+\lambda A_{\lambda}H_u\cdot H_dS+\frac{\kappa}{3}A_{\kappa} S^3+\mbox{h.c.}
                                                   \label{Lagrangian:soft}
\end{align}
From Eqs.~\eqref{Superpotential:HiggsSinglet} and \eqref{Lagrangian:soft} one obtains the Higgs scalar potential
\begin{align}
 V_{\footnotesize\mbox{Higgs}\normalsize}=\,&\left|\lambda(H^+_uH^-_d-H^0_uH^0_d)+\kappa S\right|^2\\
                                            &+(m_{H_u}^2+|\mu+\lambda S|^2)\left(|H_u^0|^2+|H_u^+|^2\right)+(m_{H_d}^2+|\mu+\lambda S|^2)\left(|H_d^0|^2+|H_d^-|^2\right)\\
                                            &+\frac{g_1^2+g_2^2}{8}\left(|H_u^0|^2+|H_u^+|^2-|H_d^0|^2-|H_d^-|^2\right)+\frac{g_2^2}{2}\left|H_u^+H_d^{0\,\ast}+H_u^0H_d^{-\,\ast}\right|^2\\
                                            &+m_S^2|S|^2+\left(\lambda A_{\lambda}(H_u^+H_d^--H_u^0H_d^0)S+\frac{k}{3}A_{\kappa} S^3+h.c.\right)\,
\end{align}
from which one derives the Higgs mass eigenstates. Conventionally, we take
\begin{equation}
 H^0_u=v_u+\frac{H_{u\,R}+iH_{u\,I}}{\sqrt{2}},\,\,\,\,\,\, H^0_d=v_d+\frac{H_{d\,R}+iH_{d\,I}}{\sqrt{2}},\,\,\,\,\,\,S=s+\frac{S_R+iS_I}{\sqrt{2}}\, .
 \end{equation}
We define $\mu_{\footnotesize\mbox{eff}\normalsize}=\lambda\, s$, so the CP-even Higgs mass matrix is given by
\begin{equation}
 \mathcal{M}^2_S=\left(\begin{array}{ccc}
                      \frac{g_1^2+g_2^2}{2}v_d^2+\mu_{\footnotesize\mbox{eff}\normalsize}(A_{\lambda}+\kappa s)\tan\beta &\left(2\lambda^2-\frac{g_1^2+g_2^2}{2}\right)v_uv_d-\mu_{\footnotesize\mbox{eff}\normalsize}(A_{\lambda}+\kappa s)&\lambda(2\mu_{\footnotesize\mbox{eff}\normalsize}v_d-(A_{\lambda}+2\kappa s)v_u)\\
                      &\frac{g_1^2+g_2^2}{2}v_u^2+\mu_{\footnotesize\mbox{eff}\normalsize}(A_{\lambda}+\kappa s)/\tan\beta&\lambda(2\mu_{\footnotesize\mbox{eff}\normalsize}v_u-(A_{\lambda}+2\kappa s)v_d)\\
                       &&\lambda A_{\lambda}\frac{v_uv_d}{s}+\kappa s
                     \end{array}
\right)\,.
\end{equation}
The CP-odd Higgs mass matrix reads:	
\begin{equation}
 \mathcal{M}^2_P=\left(\begin{array}{cc}
                      2\mu_{\footnotesize\mbox{eff}\normalsize}(A_{\lambda}+\kappa s)/\sin2\beta &\lambda(A_{\lambda}-2\kappa s)v\\
                      &\lambda(A_{\lambda}+4\kappa s)\frac{v_uv_d}{s}-3\kappa A_{\kappa}s\\
                     \end{array}
\right)\,.
\end{equation}
Finally, the NMSSM charged Higgs states $H^{\pm}$ have the mass:
\begin{equation}
 m_{H^{\pm}}^2=\frac{2\mu_{\rm eff}(A_{\lambda}+\kappa s)}{\sin2\beta}+v^2\left(\frac{g_2^2}{2}-\lambda^2\right)\,.
\end{equation}

\section{Scenarios}\label{app-c}

\subsection{Light singlino scenario}\label{Appendix:lightSinglino}Parameters at the EWSB scale (2~TeV):

\begin{center}\small
\begin{tabular}{|c|c|c|c|c|c|c|}\hline
$M_1$&$M_2$&$M_3$&tan$\beta$&$\mu_{\rm eff}=\lambda s$&$A_{\lambda}$&$A_{\kappa}$\\\hline
365 GeV&111 GeV&2000 GeV&9.5&484 GeV&4200 GeV&$-120$ GeV\\\hline
 \end{tabular} \end{center}

\begin{center}\small
\begin{tabular}{|c|c|c|c|c|c|c|}\hline
$M_{Q_{1,2}}$, $M_{u_{1,2}}$, $M_{d_{1,2}}$&$M_{Q_3}$& $M_{u_3}$& $M_{d_3}$&$M_{l}$, $M_{e}$& $A_{u_3}$&$A_{d_3}$, $A_{e_3}$\\\hline
2000 GeV&1500 GeV&1000 GeV&800 GeV&300 GeV&2750 GeV& 2000 GeV\\\hline
 \end{tabular} \end{center}

\subsection{Light higgsino scenario}\label{Appendix:lightHiggsino}Parameters at the EWSB scale (2~TeV):

\begin{center}\small
\begin{tabular}{|c|c|c|c|c|c|c|}\hline
$M_1$&$M_2$&$M_3$&tan$\beta$&$\mu_{\rm eff}=\lambda s$&$A_{\lambda}$&$A_{\kappa}$\\\hline
450 GeV&1600 GeV&2000 GeV&27&120 GeV&3000 GeV&$-30$ GeV\\\hline
 \end{tabular} \end{center}

\begin{center}\small
\begin{tabular}{|c|c|c|c|c|}\hline
$M_{Q_{1,2}}$, $M_{u_{1,2}}$,$M_{d_{1,2}}$&$M_{Q_3}$, $M_{u_3}$, $M_{d_3}$&$M_{l}$,$M_{e}$& $A_{u_3}$&$A_{d_3}$, $A_{e_3}$\\\hline
2000 GeV&1500 GeV&300 GeV&3300 GeV& 200 GeV\\\hline
 \end{tabular} \end{center}

\subsection{Light gaugino scenario}\label{Appendix:lightGaugino}Parameters at the EWSB scale (2~TeV):

\begin{center}\small
\begin{tabular}{|c|c|c|c|c|c|c|}\hline
$M_1$&$M_2$&$M_3$&tan$\beta$&$\mu_{\rm eff}=\lambda s$&$A_{\lambda}$&$A_{\kappa}$\\\hline
240 GeV&105 GeV&2000 GeV&9.2&505 GeV&3700 GeV&$-40$ GeV\\\hline
 \end{tabular} \end{center}

\begin{center}\small
\begin{tabular}{|c|c|c|c|c|c|c|c|c}\hline
$M_{Q_{1,2}}$, $M_{u_{1,2}}$,$M_{d_{1,2}}$&$M_{Q_3}$& $M_{u_3}$, $M_{d_3}$&$M_{l_{1,2}}$,$M_{e_{1,2}}$&$M_{l_3}$,$M_{e_3}$& $A_{u_3}$&$A_{d_3}$& $A_{e_3}$\\\hline
2000 GeV&1800 GeV&1500 GeV&300 GeV&500 GeV&3700 GeV& 2500 GeV&1500 GeV\\\hline
 \end{tabular} \end{center}

\bibliography{NMSSMvsMSSM_ILC}

\bibliographystyle{JHEP}

\end{document}